\documentclass{aa}

\usepackage{graphicx,aabib99,amsmath}

\newcommand{\Line}[3]{\ion{#1}{#2}\,$\lambda$\,#3}

\renewcommand{\vec}[1]{\pmb{#1}}

\begin{document}

\thesaurus{06 (02.01.2; 03.13.2; 08.02.1; 08.02.2; 08.09.2 UZ For; 08.14.2)}

\title{Eclipse Mapping of the Accretion Stream in UZ\;Fornacis\thanks{The
observational part of this work is based on observations made with the
NASA/ESA Hubble Space Telescope, obtained from the data Archive at the Space
Telescope Science Institute, which is operated by the Association of
Universities for Research in Astronomy, Inc., under NASA contract
NAS~5-26555. These observations are associated with proposal ID~4013.}}

\author{J. Kube \and B. T. G\"ansicke \and K. Beuermann}

\offprints{Jens Kube, jkube@uni-goettingen.de}

\institute{Universit\"ats-Sternwarte G\"ottingen, Geismar Landstra\ss{}e 11,
  D-37083 G\"ottingen, Germany} 

\date{Received 4 August 1999 / Accepted 20 December 1999}

\maketitle

\begin{abstract}
  
We present a new method to map the surface brightness of the accretion streams
in \object{AM Herculis} systems from observed light curves. Extensive tests of
the algorithm show that it reliably reproduces the intensity distribution of
the stream for data with a signal-to-noise ratio $\ga5$.
As a first application, we map the accretion stream emission of
\Line{C}{iv}{1550} in the polar \object{UZ Fornacis} using HST\;FOS high state
spectra.
We find three main emission regions along the accretion stream: (1) On the
ballistic part of the accretion stream, (2) on the magnetically funneled
stream near the primary accretion spot, and (3) on the magnetically funneled
stream at a position above the stagnation region.


\end{abstract}

\section{Introduction}
Polars, as AM\;Hers are commonly named due to their highly polarized emission,
consist of a late type main-sequence star (red dwarf, secondary star) and a
highly magnetized white dwarf (WD, primary star).
The red dwarf, filling its Roche volume, injects matter through the $L_1$
point into the Roche volume of the WD.  Unlike in non-magnetic systems, this
material does not form an accretion disc, but couples onto the magnetic field
once the magnetic pressure exceeds the ram pressure. From this stagnation
region ($SR$) on, the accretion stream follows a magnetic field line until it
impacts onto the white dwarf surface. For a review of polars, see Warner
\cite*{Warner95Polars}.

In systems with an inclination $i\ga70\degr$, the secondary star
gradually eclipses the accretion stream during the inferior
conjunction.  Using tomographical methods, it is -- in principle --
possible to reconstruct the surface brightness distribution on the
accretion stream from time resolved observations.  This method has
been successfully applied to accretion discs in non-magnetic CVs
(``eclipse mapping'', Horne \cite*{H85}).  We present tests and a
first application of a new eclipse mapping code, which allows the
reconstruction of the intensity distribution on the accretion stream
in magnetic CVs.

\begin{figure}
\includegraphics[width=8.8cm]{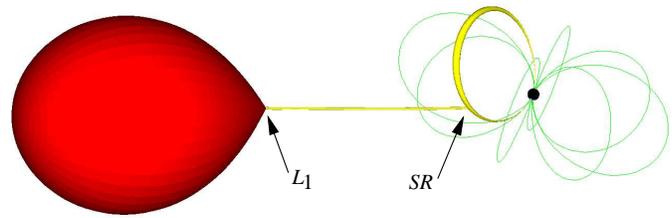}
\caption{Overview of our 3d-Model of a polar. $L_1$ marks the inner
lagrangian point, $SR$ the position of the stagnation or coupling
region.}
\label{f_overview}
\end{figure}

Similar attempts to map accretion streams in polars have been investigated by
Hakala \cite*{H95} and Vrielmann and Schwope \cite*{VS98} for
\object{HU~Aquarii}. An improved version of Hakala's \cite*{H95} method has
been presented by Harrop-Allin et al. \cite*{HHC99}
with application to real data for the system
\object{HU\;Aquarii} \cite{HCH99}. A drawback of all these approaches is
that they only consider the eclipse of the accretion stream by the secondary
star. In reality, the geometry may be more complicated: the far side of the
magnetically coupled stream may eclipse stream elements close to the WD, as
well as the hot accretion spot on the WD itself. The latter effect is commonly
observed as a dip in the soft X-ray light curves prior to the eclipse
(e.g. Sirk and Howell, 1998\nocite{SH98}). The stream-stream eclipse may be
detected in data which are dominated by emission from the accretion stream,
e.g. in the light curves of high-excitation emission lines where the secondary
contributes only little to the line flux.

Here, we describe a new accretion stream eclipse mapping method, using a 3d
code which can handle the full complexity of the geometry together with an
evolution strategy as fit algorithm. We present extensive tests of the method
and map as a first application to real data the accretion stream in UZ\;For
emission of \Line{C}{iv}{1550}. 

\section{The 3d Cataclysmic Variable Model\label{s_Method}}

Our computer code CVMOD generates $N$ small surface elements (convex
quadrangles, some of which are degenerated to triangles), which represent the
surfaces of the individual components of the CV (WD, secondary, accretion
stream) in three-dimensional space (Fig.~\ref{f_overview}). Using simple
rotation algorithms, the position of each surface element $i=1\dots N$ can be
computed for a given orbital phase $\Phi$.

The \emph{white dwarf} is modelled as an approximated sphere, using surface
elements of nearly constant area \cite{GHB98}. The \emph{secondary star} is
assumed to fill its Roche volume. Here, the surface elements are choosen in
such a way that their boundaries align with longitude and latitude circles of
the Roche surface, taking the $L_1$-point as the origin.

We generate the surface of the \emph{accretion stream} in two parts, (a) the
ballistic part from $L_1$ to $SR$, and (b) the dipole-part from $SR$ to the
surface of the white dwarf.

(a) For the ballistic part of the stream, we use single-particle
trajectories. The equations of motion in the corotating frame are
given by
\begin{eqnarray}
\ddot x&=&+\mu\frac{x-x_1}{r_1^3}-(1-\mu)\frac{x-x_2}{r_2^3}+2\dot
y+x\label{m_method_2a}\\ \ddot
y&=&-\mu\frac{y}{r_1^3}-(1-\mu)\frac{y}{r_2^3}-2\dot x+y\label{m_method_2b}\\
\ddot z&=&-\mu\frac{z}{r_1^3}-(1-\mu)\frac{z}{r_2^3}\label{m_method_2c}
\end{eqnarray} 
Equation (\ref{m_method_2c}) has been added to Flannery's \cite*{F75a} set of
two-dimensional equations. $\mu=(M_1+M_2)/M_1$ is the mass fraction of the
white dwarf, $r_1$ and $r_2$ are the distances from the point $(x,y,z)$ to the
white dwarf and the secondary, respectively, in units of the orbital
separation $a$. The coordinate origin is at the centre of gravity, the
$x$-axis is along the lines connecting the centres of the stars, the system
rotates with the angular frequency $\omega$ around to the $z$-axis. The
velocity $\vec{v}=(\dot x,\dot y,\dot z)$ is given in units of $a\omega$,
$\vec{v_0}=(\dot x_0,\dot y_0,\dot z_0)$ is the initial velocity in the $L_1$
point.

If $\dot z_{0}=0$, the trajectories resulting from the numerical integration
of eq. (\ref{m_method_2a}) -- (\ref{m_method_2c}) are restricted to the
orbital plane. However, calculating single-particle trajectories with
different initial velocity directions (allowing also $\dot z_0\neq0$) shows
that there is a region approximately one third of the way downstream from
$L_1$ to $SR$ where all trajectories pass within very small separations,
corresponding to a striction of the accretion stream.

We define a 3d version of the stream as a tube with a circular cross section
with radius $r_{\mathrm{Tube}}=5\times 10^8\,\mathrm{cm}$ centred on the
single-particle trajectory for $\dot x=10\,\mathrm{km}\,\mathrm{s}^{-1}, \dot
y=\dot z=0$.

(b) When the matter reaches $SR$, we switch from a ballistic
single-particle trajectory to a magnetically forced dipole geometry. The
central trajectory is generated using the dipole formula $r=\pmb{r_0}\sin^2\alpha$,
where $\alpha$ is the angle between the dipole axis and the position
of the particle $(r,\varphi,\alpha)$.  This can be interpreted as the
magnetic field line $F$ passing through the stagnation point $SR$ and the hot
spots on the WD.
Knowing $F$, we assume a circular cross section with the radius
$r_{SR}=r_{\mathrm{Tube}}=5\times10^8\,\mathrm{cm}$ for the region where the
dipole intersects the ballistic stream. This cross section is subject to
transformation as $\alpha$ changes. Thus, the cross section of the stream is
no longer constant in space but bounded everywhere by the same magnetic field
lines.

Our accretion stream model involves several assumptions: (1) The cross section
of the stream itself is to some extent arbitrary because we consider it to be
-- for our current data, see below -- essentially a line source. (2) The
neglect of the magnetic drag \cite{K93,WK95} on the ballistic part of the
stream and the neglect of deformation of the dipole field cause the model
stream to deviate in space from the true stream trajectory. While, in fact,
the location of $SR$ may fluctuate with accretion rate (as does the location
of the Earth's magnetopause), the evidence for a sharp soft X-ray absorption
dip caused by the stream suggest that $SR$ does not wander about on time
scales short compared to the orbital period. (3) The
abrupt switch-over from the ballistic to the dipole part of the stream may not
describe the physics of $SR$ correctly. This discrepancy, however, does not
seriously affect our results, because the eclipse tomography is sensitive
primarily to displacements in the times of ingress and egress of $SR$ which
are constrained by the absorption dip in the UV continuum (and, in principle,
in soft X-rays). The $\approx 5\,\mbox{sec}$ time bins of our observed light
curves correspond to $\approx 10^8\,\mbox{cm}$ in space at $SR$. Hence, our
code is insensitive to structure on a smaller scale. In fact, the smallest
resolved structures are much larger because of the noise level of our
data. While our approach clearly involves several approximations, it is
tailored to the desired aim of mapping the accretion stream from the
information obtained from an emission line light curve.

In our current code, we restrict the possible brightness distribution so that
for each stream segment, which consists of 16 surface elements forming a
section of the tube-like stream, the intensity is the same, i.e. there is no
intensity variation around the stream.  For the current data, this is
no serious drawback, since we only use observations covering a small phase
interval around the eclipse. Our results refer, therefore, to the stream
brightness as seen from the secondary. From the present
observations we can not infer how the fraction of the stream illuminated by
the X-ray/UV spot on the WD looks like. The required extension of our computer
code, allowing for brightness variation around the stream, is
straightforward. The present version of the code is, however, adapted to the
data set considered here.

\section{Light curve fitting}

The basic idea of our eclipse mapping algorithm is to reconstruct the
intensity distribution on the accretion stream by comparing and fitting a
synthetic light curve to an observed one. The comparison between these light
curves is done with a $\chi^2$-minimization, which is modified by means of a
maximum entropy method. Sections \ref{lcg} describes the light curve
generation, \ref{ctp} the maximum entropy method, and \ref{tfa} the actual
fitting algorithm.

\subsection{Light curve generation}

\label{lcg}

In order to generate a light curve from the 3d model, it is neccessary to know
which surface elements $i$ are visible at a given phase $\Phi$. We designate
the \emph{set of visible surfaces} $V(\Phi)$.

In general, each of the three components (WD, secondary, accretion stream) may
eclipse (parts of) the other two, and the accretion stream may partially
eclipse itself.
This is a typical hidden surface problem. However, in contrast to the
widespread computer graphics algorithms which work in the image space
of the selected output device (e.g. a screen or a printer), and which
provide the information `pixel $j$ shows surface $i$', we need to work
in the object space, answering the question `is surface $i$ visible at
phase $\Phi$?'. 
For a recent review on object space algorithms see Dorward
\cite*{D94}. Unfortunately, there is no readily available algorithm which fits
our needs, thus we use a self-designed 3d object-space hidden-surface
algorithm. Let $N$ be the number of surface elements of our 3d
model. According to Dorward \cite*{D94}, the time $T$ needed to perform an
object space visibility analysis goes like $T\propto N\log N\dots N^2$. Our
algorithm performs its task in $T\propto N^{1.5\dots1.8}$, with the faster
results during the eclipse of the system. It is obviously necessary to
optimize the number of surface elements in order to minimize the computation
time without getting too coarse a 3d grid.

Once $V(\Phi)$ has been determined, the angles between the surface normals of
$i\in V(\Phi)$ and the line of sight, and the projected areas $A_{i,\Phi}$ of
$i\in V(\Phi)$ are computed.  Designating the intensity of the surface element
$i$ at the wavelength $\lambda$ with $I_{i,\lambda}$, the observed flux
$F_\lambda(\Phi)$ is
\begin{equation}
  F_\lambda(\Phi)=\sum\limits_{i\in
  V(\Phi)}I_{i,\lambda}A_i(\Phi)\label{e_flux}
\end{equation}
Here, two important assumptions are made: (a) the emission from all
surface elements is optically thick, and (b) the emission is
isotropic, i.e. there is no limb darkening in addition to the
foreshortening of the projected area of the surface
elements. 
The computation of a synthetic light curve is straightforward. It suffices
to compute $F_\lambda(\Phi)$ for the desired set of orbital phases.

While the above mentioned algorithm can produce light curves for all three
components, the WD, the secondary, and the accretion stream, we constrain in
the following the computations of light curves to emission from the accretion
stream only. Therefore, we treat the white dwarf and the secondary star as
dark opaque objects, screening the accretion stream.

\subsection{Constraining the problem: MEM}

\label{ctp}

In the eclipse mapping analysis, the number of free parameters, i.e. the
intensity of the $N$ surface elements, is typically much larger than the
number of observed data points. Therefore, one has to reduce the degrees of
freedom in the fit algorithm in a sensible way.
An approach which has proved successful for accretion discs is the maximum
entropy method MEM \cite{H85}. The basic idea is to define an image entropy
$S$ which has to be maximized, while the deviation between synthetic and
observed light curve, usually measured by $\chi^2/n$, is minimized ($n$ is the
number of phase steps or data points). Let $D_i$ be
\begin{equation}
\label{e_mem-default}
D_i=\frac{\sum\limits_{j=1}^{N}I_j\exp\left(-\frac{\displaystyle(\vec{r}_i-\vec{r}_j)^2}{\displaystyle2\Delta^2}\right)}{\sum\limits_{j=1}^{N}\exp\left(-\frac{\displaystyle(\vec{r}_i-\vec{r}_j)^2}{\displaystyle2\Delta^2}\right)}
\end{equation}
the default image for the surface element $i$. Then the entropy is given by
\begin{equation}
\label{e_mem-entropy}
S=\frac{\sum\limits_{i=1}^{N}I_i\left(\ln{\displaystyle\frac{I_i}{D_i}}-1\right)}{\sum\limits_{i=1}^{N}I_i}
\end{equation}
In Eq. (\ref{e_mem-default}), $\vec{r}_i$ and $\vec{r}_j$ are the positions of
the surface elements $i$ and $j$. $\Delta$ determines the range of the MEM in
that the default image (\ref{e_mem-default}) is a convolution of the actual
image with a Gaussian with the $\sigma$-width of $\Delta$.  Hence, the entropy
measures the deviation of the actual image from the default image.  An ideal
entropic image (with no contrast at all) has $S=1$. We use
$\Delta=1\times10^9\,\mbox{cm}\approx 0.02a$ for our test calculations and for
the application to UZ\;For.

The quality of a intensity map is given as 
\begin{equation}
\label{e_quality}
{\cal Q}=\chi^{2}/n-\lambda S,
\end{equation}
where $\lambda$ is chosen in the order of 1. Aim of the fit algorithm
is to minimize ${\cal Q}$.

\subsection{The fitting algorithm: Evolution strategy\label{s_evo}}

\label{tfa}

Our model involves approximately 250 parameters, which are the intensities of
the surface elements. This large number is \emph{not} the number
of the degrees of freedom, which is difficult to define in a MEM-strategy. A
suitable method to find a parameter optimum with a least $\chi^2$ and a
maximum entropy value is a simplified imitiation of biological evolution,
commonly referred to as `evolution strategy' \cite{Rechenberg94}. The
intensity information of the $i$ surface elements is stored in the intensity
vector $\vec{I}$. Initially, we choose $I_i=1$ for all $i$.

>From this parent intensity map, a number of offsprings are created
with $I_i$ randomly changed by a small amount, the so-called {\em
mutation\/}.  For all offsprings, the quality ${\cal Q}$ is
calculated. The best offspring is selected to be the parent of the
next generation. An important feature of the evolution strategy is
that the amount of mutation itself is also being evolved just as if it
were part of the parameter vector.
We use the C-program library evoC developed by K. Trint and U. Utecht from the
Technische Universit\"at Berlin, which handles all the important steps
(offspring generation, selection, stepwidth control).

In contrast to the classical maximum entropy optimisation
\cite{SB84}, the evolution strategy does not
offer a quality parameter that indicates how close the best-fit
solution is to the global optimum. In order to test the stability of
our method, we run the fit several times starting from randomly
distributed maps.  All runs converge to very similar intensity
distributions $\vec{I}$ (see also Figs. \ref{f_lcs} and
\ref{f_result_map_plot}). This type of test is common in evolution
strategy or genetic algorithms (e.g. Hakala 1995).\nocite{H95} Even
though this approach is not a statistically `clean' test, it leaves us
to conclude that we find the global optimum.

Fastest convergence is achieved with 40 to 100 offsprings in each
generation. 
Finding a good fit ($\chi^2/n$) takes only on tenth to one fifth of the total
computation time, the remaining iterations are needed to improve the
smoothness of the intensity map, i.e. to maximize $S$.  A hybrid code using a
classical optimization algorithm, e.g. Powell's method, may speed up the
regularization \cite{PHC98}.

\begin{table}
\caption{System geometry of the imaginary system IM\;Sys}
\label{t_test-geometry}
\begin{flushleft}
\begin{tabular}{ll}
\hline
mass ratio&$Q=M_1/M_2=4$\\
total mass&$M=M_1+M_2=0.9M_{\sun}$\\
orbital period&$P=100\,\mbox{min}$\\
inclination&$i=88\degr$\\
dipole tilt&$\beta=20\degr$\\
dipole azimuth&$\Psi=35\degr$\\
angle to $SR$&$\Psi_S=35\degr$\\
\hline
\end{tabular}
\end{flushleft}
\end{table}

\begin{figure*}
\begin{minipage}{8.8cm}
\begin{center}
\includegraphics[width=7cm]{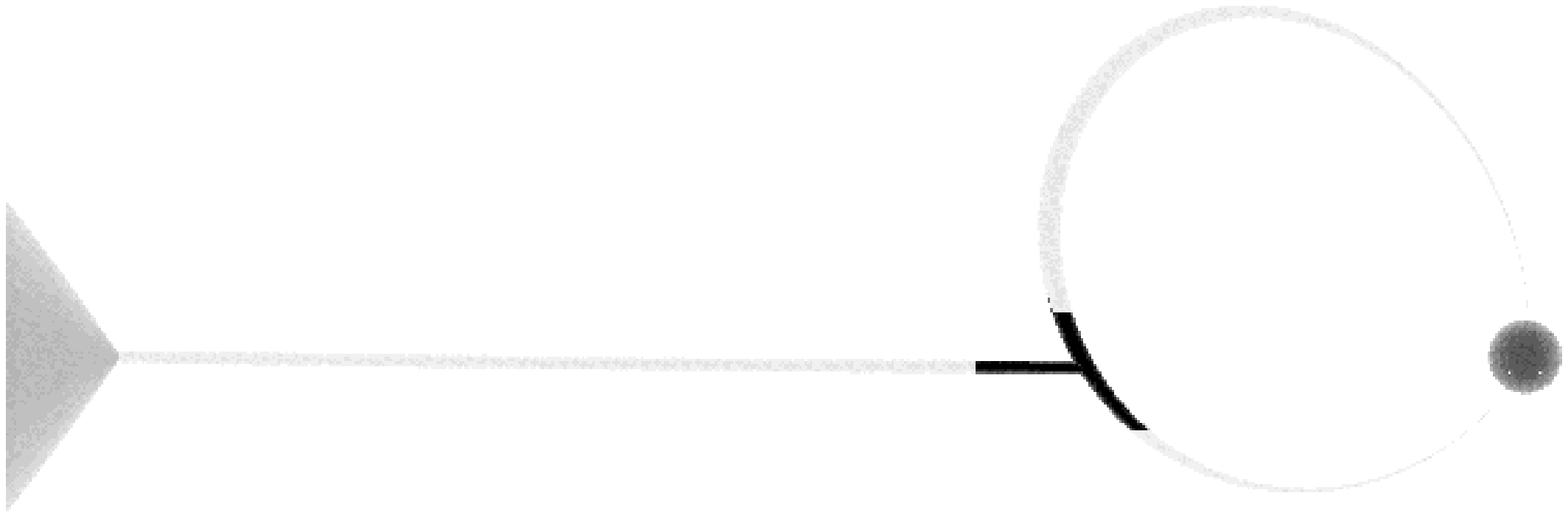}\\
\includegraphics[width=7cm]{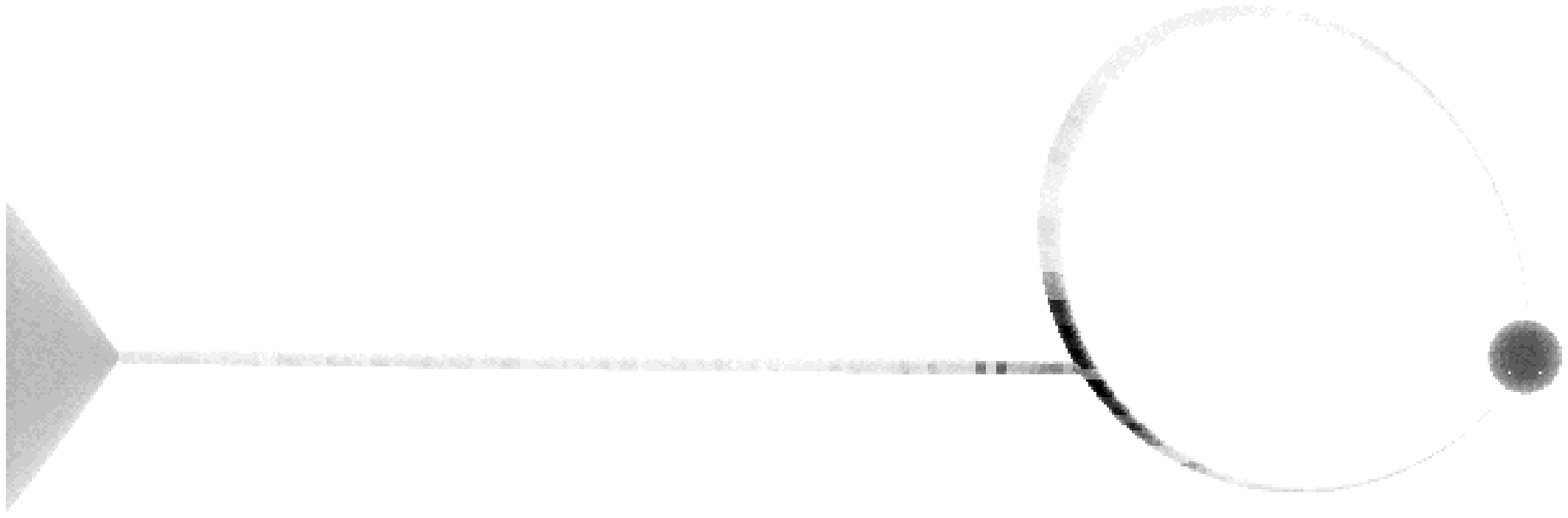}\\
\includegraphics[width=7cm]{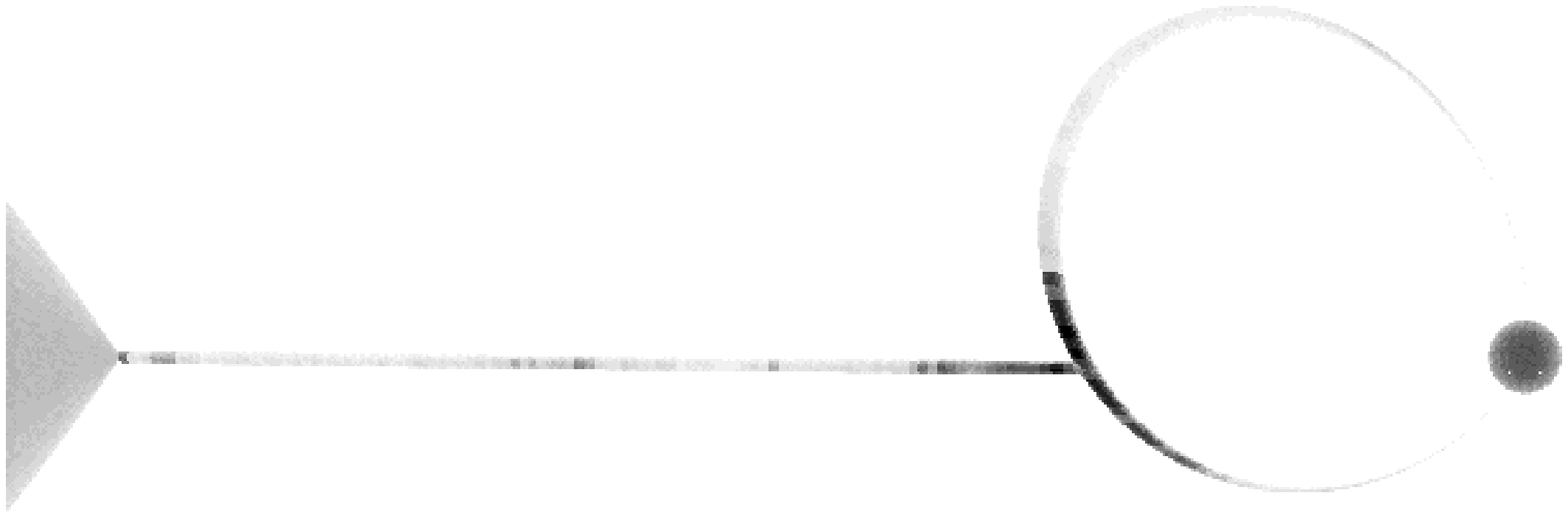}\\
\includegraphics[width=7cm]{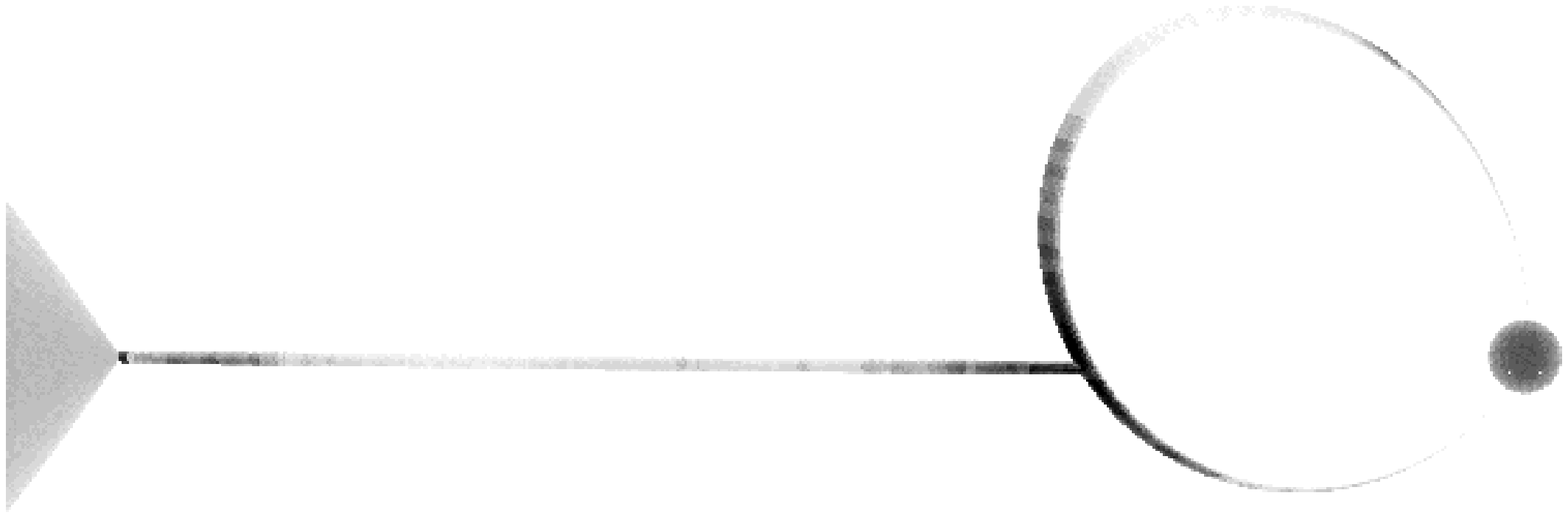}
\end{center}
\end{minipage}
\hfill
\begin{minipage}{8.8cm}
\begin{center}
\includegraphics[width=8.8cm]{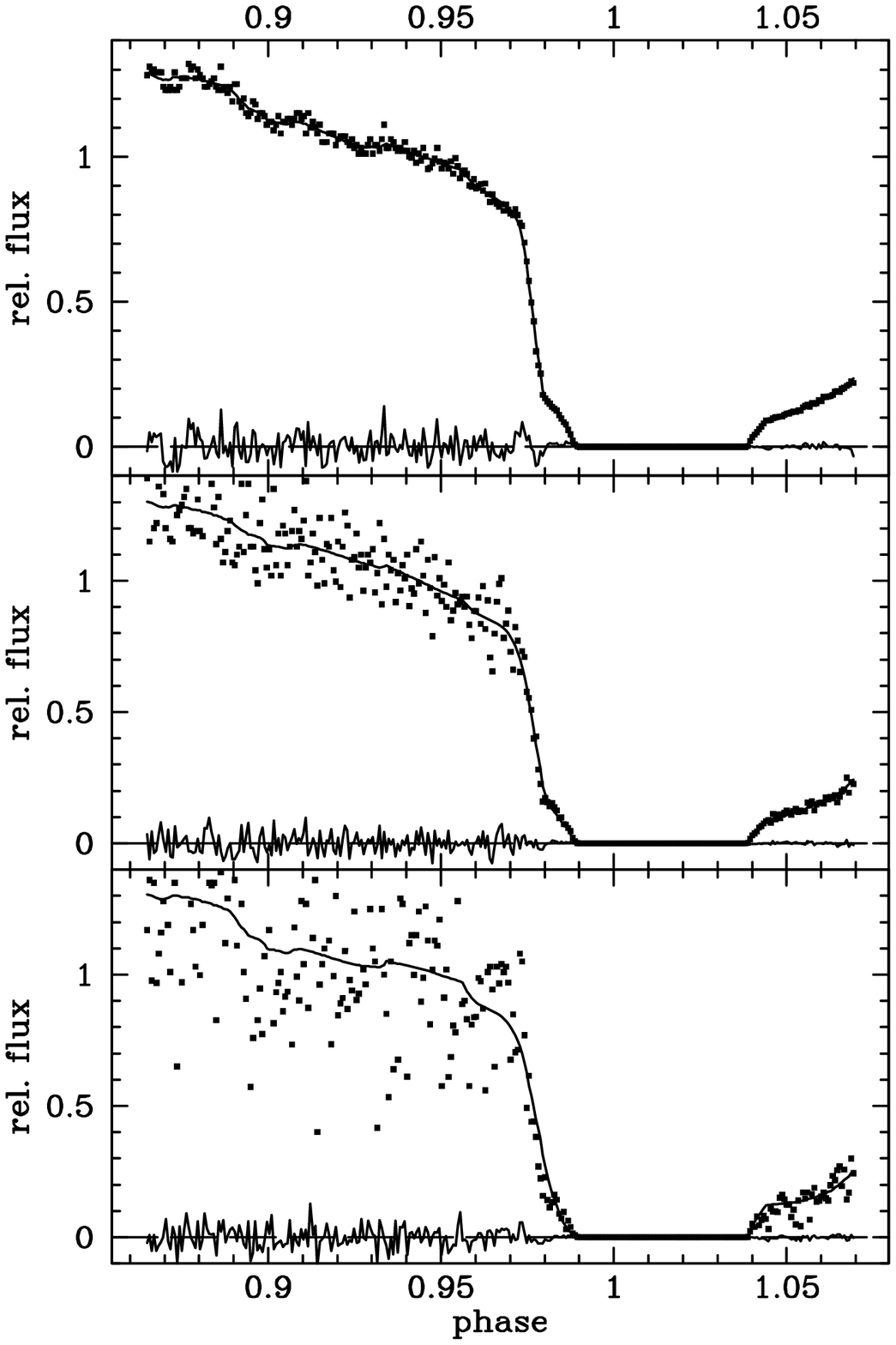}
\end{center}
\end{minipage}

\parbox[t]{8.8cm}{
\caption{\label{f_map_point} Maps of the synthetic stream and its
reconstructions. From top to bottom: Input data, reconstructions with
$\mbox{\mbox{S/N}}=50,10,4$.}}
\hfill
\parbox[t]{8.8cm}{
\caption{\label{f_test_point} Synthetic light curve of an accretion stream
which is bright around the stagnation region. Different levels of artifical
noise are added: $\mbox{\mbox{S/N}}=50,10,4$. The reconstructed light curves
are shown as solid lines. The residuals are normalized so that the standard
deviation $\sigma$ is 0.1 in the relative flux units.}}

\parbox{18cm}{
\bigskip
\parbox{11cm}{\includegraphics[width=11cm]{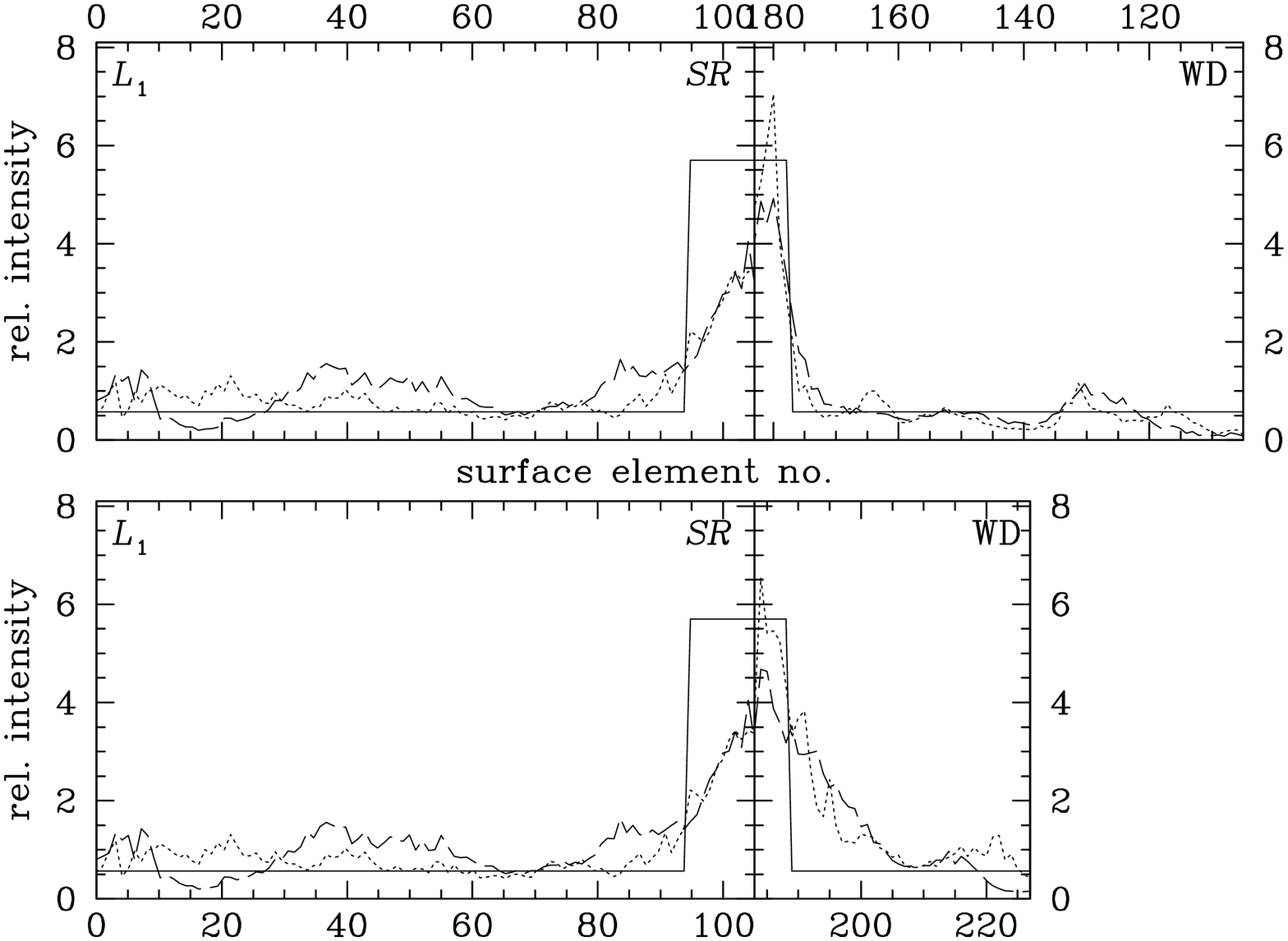}}
\hfill
\parbox{6.8cm}{
\caption{\label{f_plot_point} Plot of the reconstruction of the synthetic
stream with a bright region around $SR$. Solid line: Input distribution,
dotted line: reconstruction with $\mbox{\mbox{S/N}}=50$, dashed line:
reconstruction with $\mbox{\mbox{S/N}}=10$.  In the left panels (surface
elements no. 0 to 104), the intensities of the surface elements on the
ballistic stream are shown. In the upper right panel, the northern magnetic
stream is to be found (surface elements no. 105 to 182), in the lower right
panel the southern magnetic stream (surface elements no. 183 to 227).}}
}
\end{figure*}

\section{Tests}
To test the quality and the limits of our method, we produce synthetical test
light curves with different noise levels
($\mbox{\mbox{S/N}}=\infty,50,20,10,4$). We then try to reproduce our initial
intensity distribution on the stream from the synthetic data. Two tests with
different intensity distributions are performed. For both tests, the geometry
of the imaginary system IM\;Sys is chosen as shown in Table
\ref{t_test-geometry}. The phase coverage is $\Phi=0.865\dots1.070$ with 308
equidistant steps, which is identical to the real HST data of UZ\;For which
we use below for a first application.

Additionally, we test our algorithm with a full-orbit light curve
with $\mbox{S/N}=10$ to demonstrate its capabilities if more than just the
ingress information for each stream section is available.

\subsection{One bright region near $SR$}

\label{ObrnSR}

In the first test, we set the stream brightness to 1 on the whole stream
except for a small region near the stagnation point $SR$, where the intensity
is set to 10. For this intensity distribution, we show the theoretical light
curve in Fig.~\ref{f_test_point}. One clearly sees the fast ingress of the
small bright region at phase $\Phi=0.975$, whereas the egress occurs beyond
$\Phi=1.07$ and is not covered as in the real data of UZ\;For.

The initial intensity map and the reconstructed map with the different
noise levels are shown in Figs. \ref{f_map_point} and
\ref{f_plot_point}. The reconstruction of one bright region near the
stagnation point is achieved with no artifacts, not too much smearing
and little noise for $\mbox{S/N}\geq10$. Even with $\mbox{S/N}=4$, a
reasonable reconstruction can be obtained, but with artifacts: The
ballistic stream appears bright near the $L_1$-point, and an
additional bright region appears on the dipole stream near the
northern accretion pole.

\subsection{Three bright regions on the ballistic stream}
\label{Tbrotbs}

\begin{figure*}

\begin{minipage}{8.8cm}
\begin{center}
\includegraphics[width=7cm]{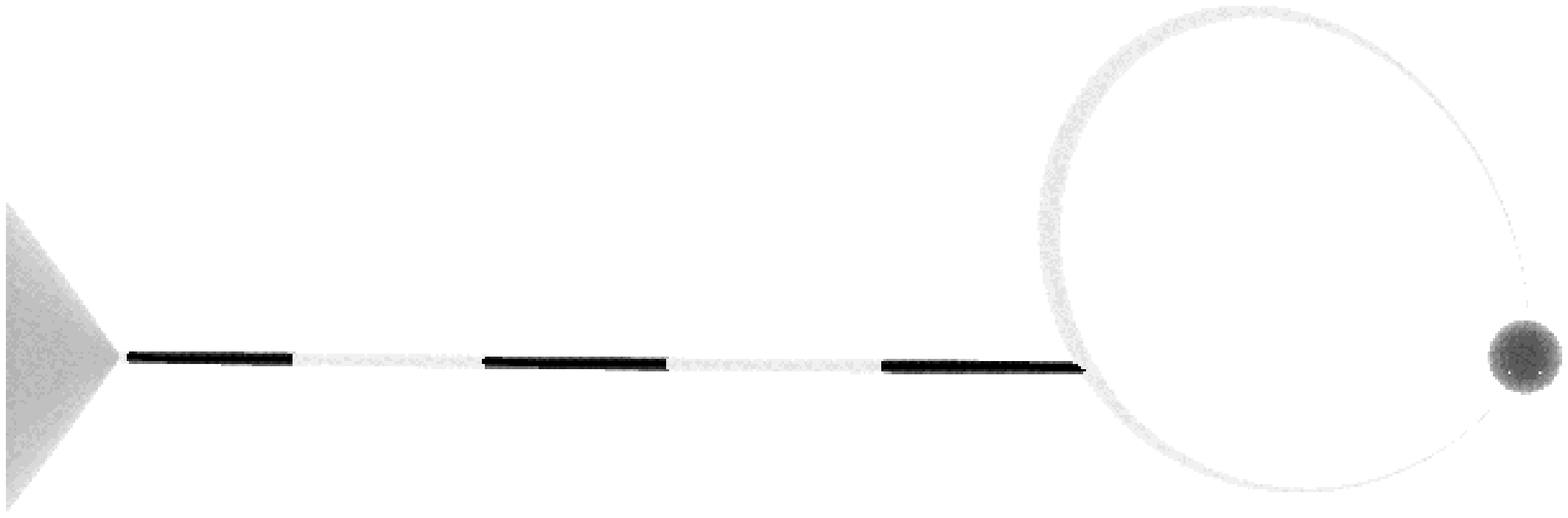}\\
\includegraphics[width=7cm]{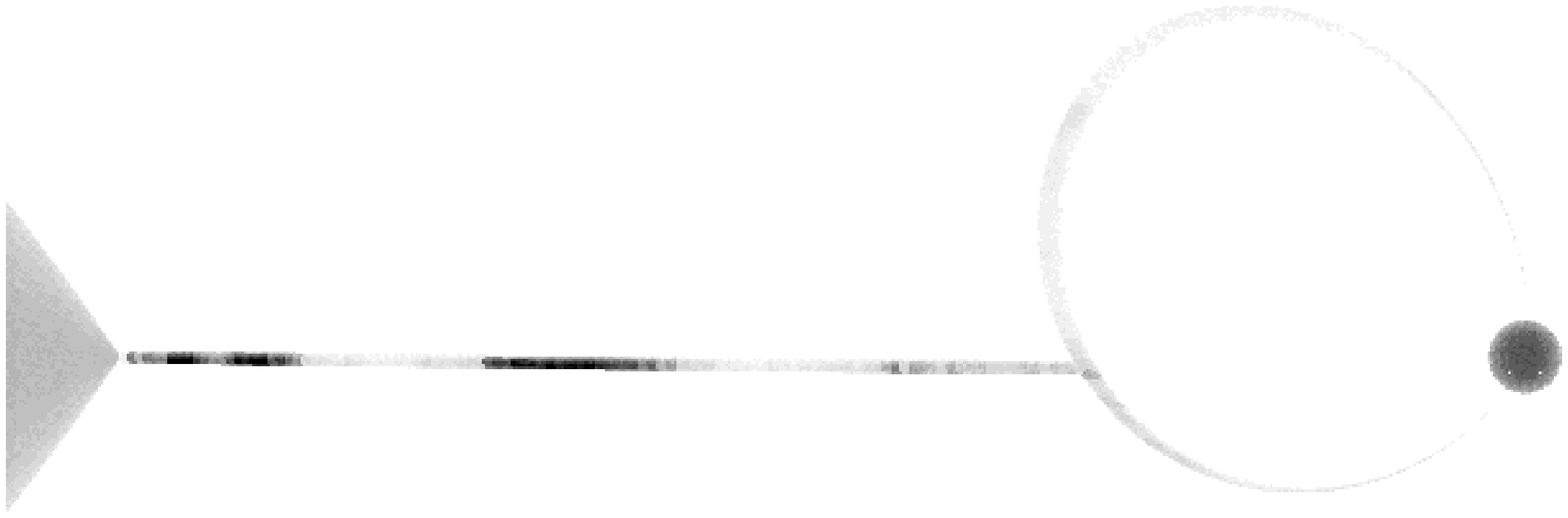}\\
\includegraphics[width=7cm]{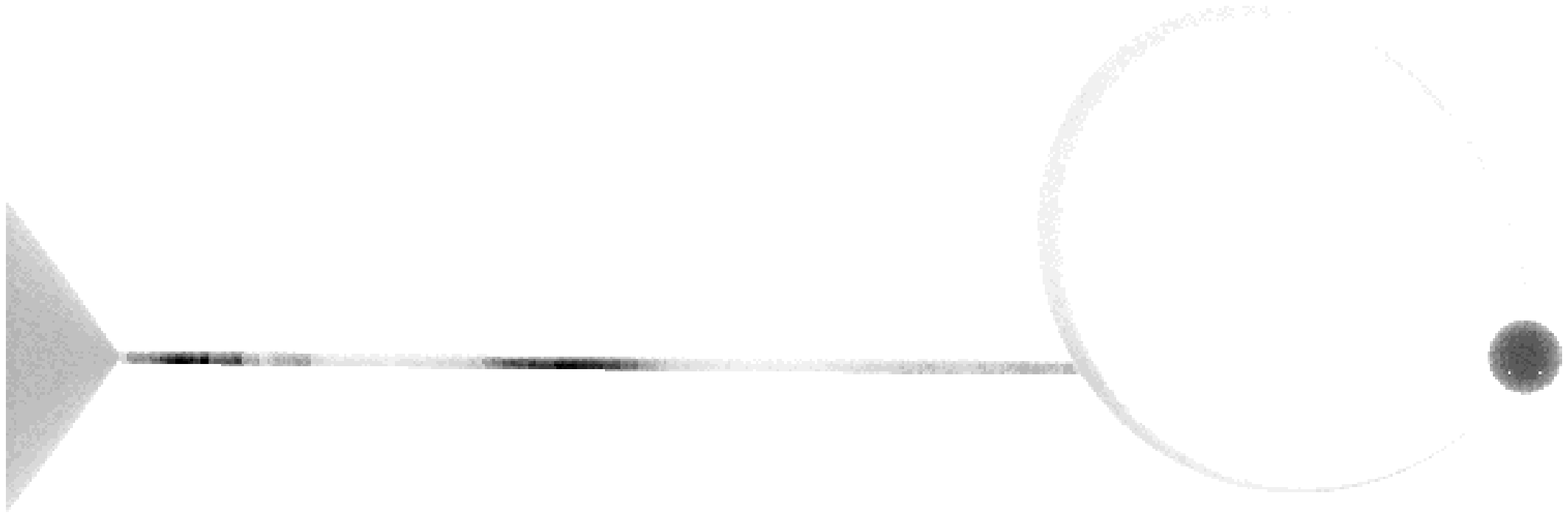}\\
\includegraphics[width=7cm]{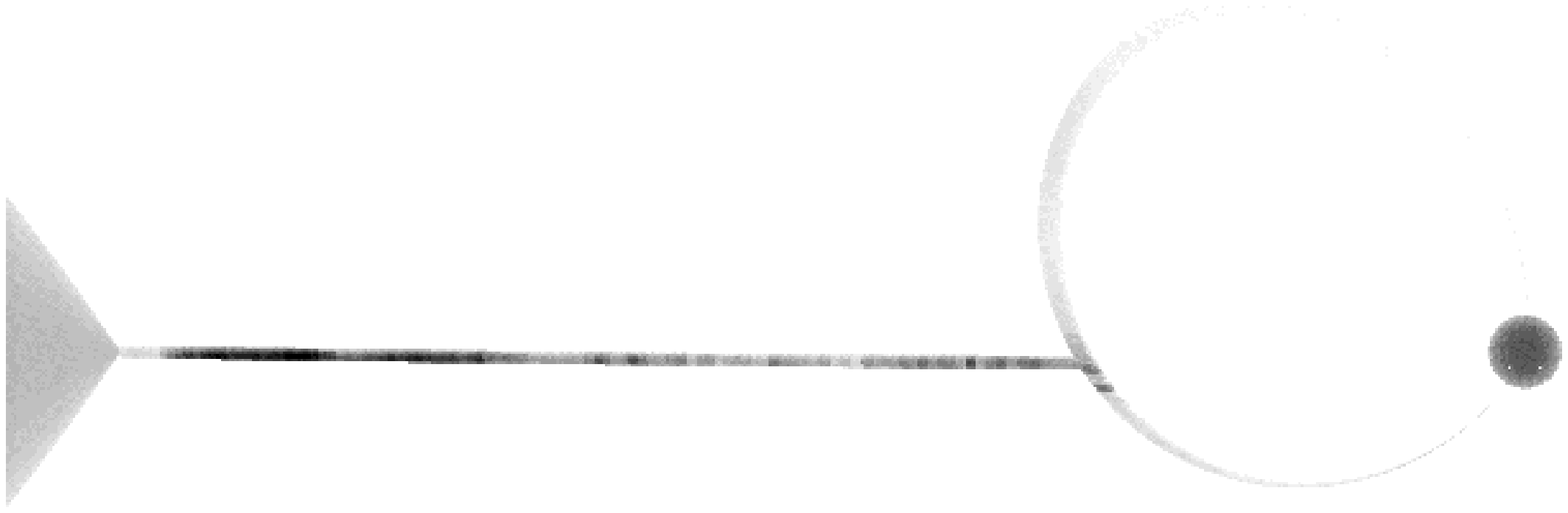}
\end{center}
\end{minipage}
\hfill
\begin{minipage}{8.8cm}
\begin{center}
\includegraphics[width=8.8cm]{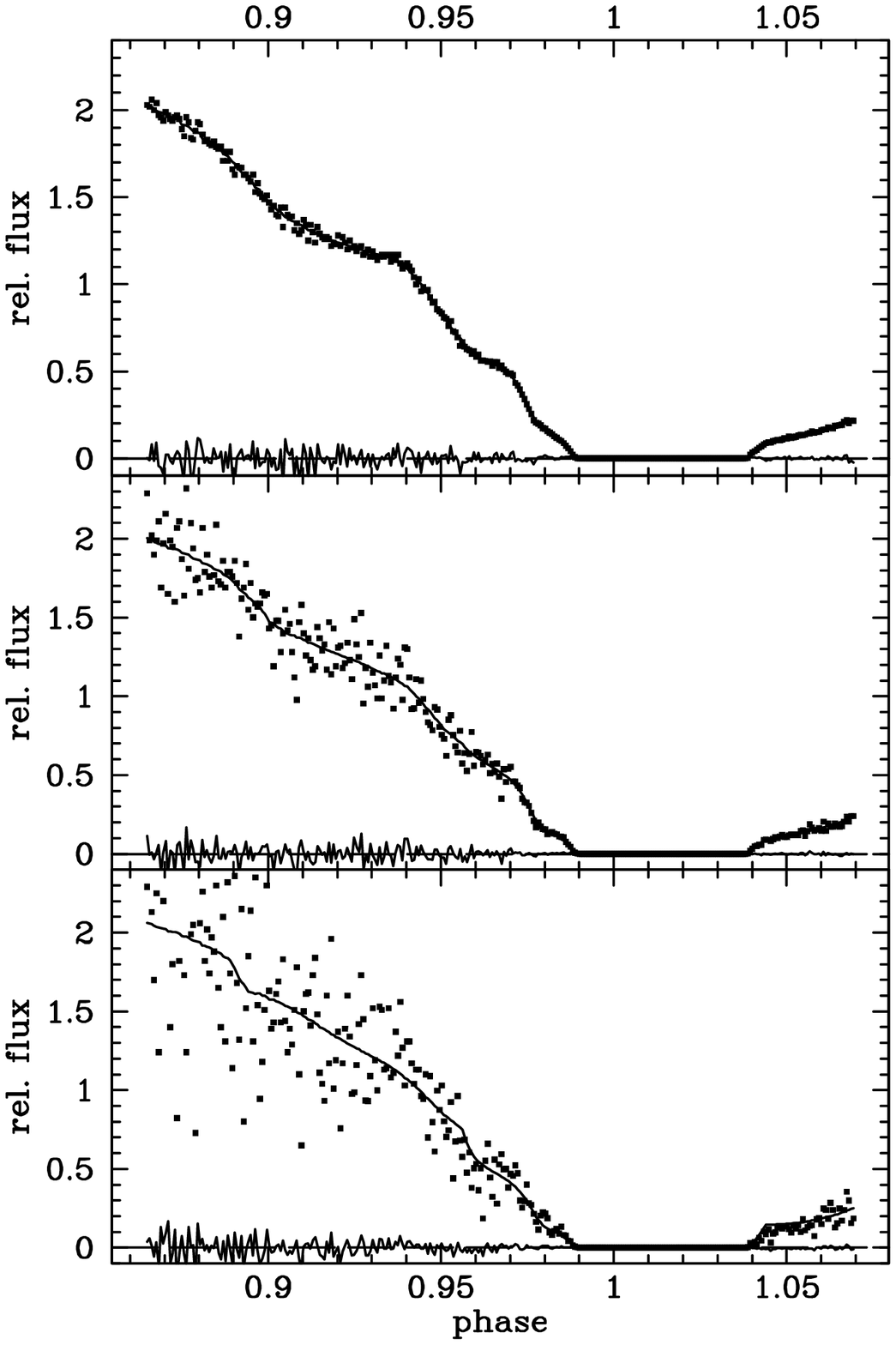}
\end{center}
\end{minipage}

\parbox[t]{8.8cm}{
\caption{\label{f_map_zebra}Maps of the synthetic stream and its
reconstructions. From top to bottom: Input data, reconstructions with
$\mbox{\mbox{S/N}}=50,10,4$.}}
\hfill
\parbox[t]{8.8cm}{
\caption{\label{f_test_zebra}Synthetic light curve of an accretion stream with
three bright parts on the ballistic stream. Different levels of artifical
noise are added: $\mbox{\mbox{S/N}}=50,10,4$. The reconstructed light curves
are shown as solid lines. The residuals are normalized so that the standard
deviation $\sigma$ is at 0.1 in the relative flux.}}

\parbox{18cm}{
\bigskip
\parbox{11cm}{\includegraphics[width=11cm]{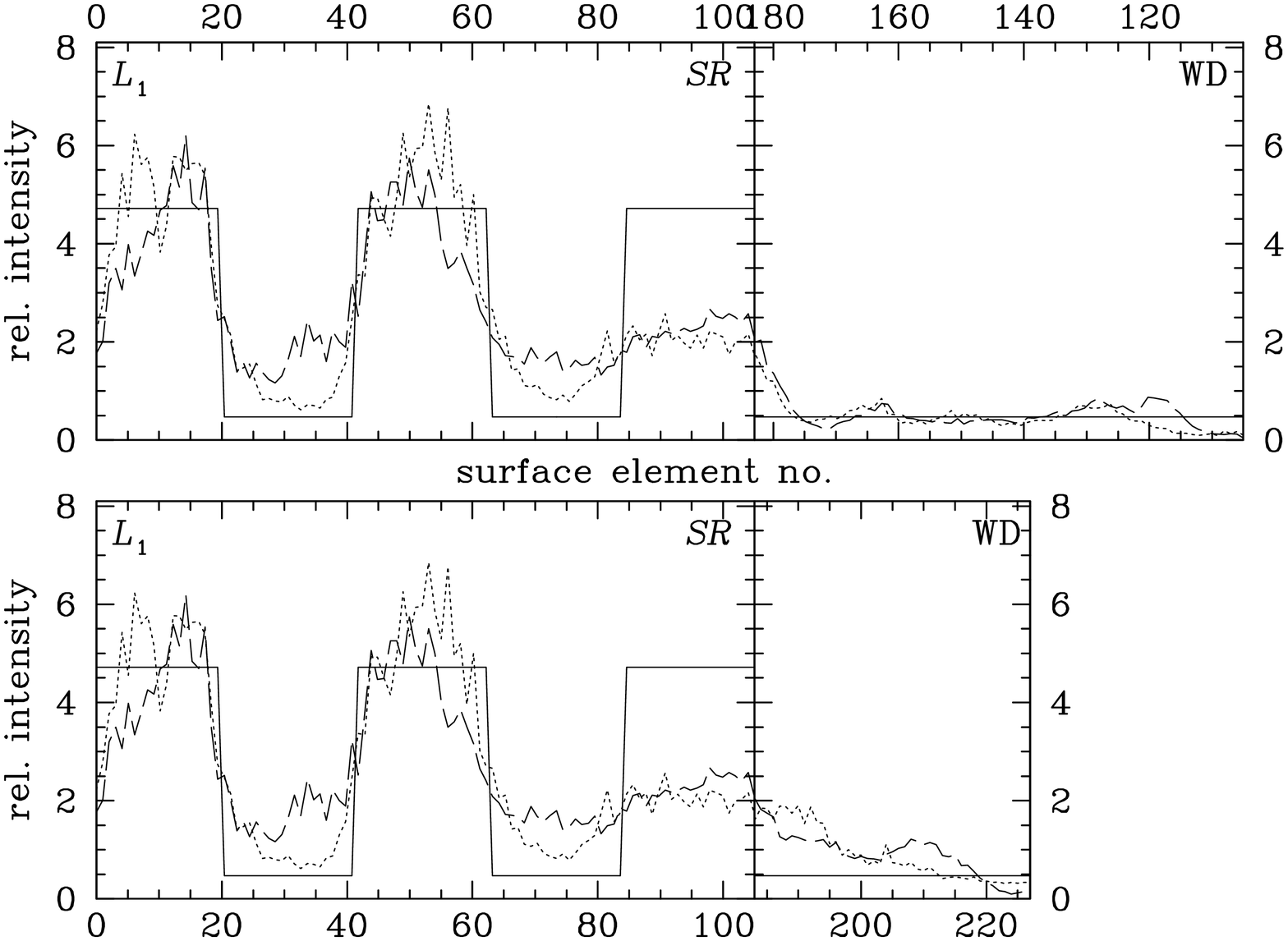}}
\hfill
\parbox{6.5cm}{
\caption{\label{f_plot_zebra}Plot of the reconstruction of the
synthetic stream with three bright regions on the ballistic
stream. Solid line: Input distribution, dotted line: reconstruction
with $\mbox{\mbox{S/N}}=50$, dashed line: reconstruction with
$\mbox{\mbox{S/N}}=10$. For an explanation of the plot see
Fig.~\ref{f_plot_point}}}
}
\end{figure*}

In the second test we assume a rather unphysical intensity distribution with
the aim to test the spatial resolution of our mapping method: The ballistic
part of the accretion stream between $L_1$ and $SR$ is divided into 5 sections
of equal length. Alternately, the intensity on these sections is set equal to
10 and to 1, producing a `zebra'-like pattern. The intensity on the
magnetically funneled stream is set to $I=1$. The synthetic light curve for
this intensity distribution differs strongly from that in out first test
(Fig.~\ref{f_test_zebra}). Instead of one sharp step in the light curve, there
are now -- as expected -- three steps during the ingress and three during the
egress (also not visible in the selected phase interval).

\begin{table}
\caption{Quality and number of iterations
for the test calculations.}
\label{t_test_quality}
\begin{flushleft}
\begin{tabular}{ccccc}
\hline
$\mbox{\mbox{S/N}}$&$\chi^2/n$&$S$&$\chi^2/n$&$S$\\
\hline
&\multicolumn{2}{c}{\emph{bright region near $SR$}}&\multicolumn{2}{c}{\emph{three bright regions}}\\
$\infty$&0.015&0.982&0.018&0.938\\
50&0.173&0.947&0.284&0.960\\
20&0.593&0.968&0.668&0.953\\
10&0.886&0.971&0.924&0.983\\
 4&1.038&0.980&1.038&0.984\\
\hline
\end{tabular}
\end{flushleft}
\end{table}

In Figs.~\ref{f_map_zebra} and \ref{f_plot_zebra} we show the input and the
resulting maps for the `zebra'-test. As long as $\mbox{\mbox{S/N}}\geq 10$,
our algorithm achieves, as in the first test, a good reconstruction of the
input map. The bright regions next to $SR$ appear darker than in the original
map, since the intensity from that region is spread over the last part of the
ballistic stream and the neighbouring parts of the dipole stream, which
dissapear behind the limb of the secondary star nearly simultaneously during
ingress.

Comparing the two tests for $\mbox{\mbox{S/N}}=4$, we find that it is still
possible to distinguish between the two different initial intensity
distributions (one bright region and `zebra') even with such a noisy signal.

In Table \ref{t_test_quality}, we list the quality parameters of the fits. The
increase of $\chi^2/n$ from $\approx 0$ to $\approx 1$ with the decrease of
$\mbox{\mbox{S/N}}$ shows once more how well the fit algorithm works: For the
light curve with no noise, the ideal light curve has -- by definition --
$\chi^2=0$. As soon as noise is added, even the input light curve has a
$\chi^2/n\approx1$, which follows also from the definition of $\chi^2$ if
$\sigma$ is the real standard deviation of the data set. For the calculation
of $\chi^2$, one needs a sensibly chosen standard deviation $\sigma$. Since
the light curves of polars show pronounced flickering, we calculate an
approximate $\sigma$ for our noisy data sets (synthetic and measured) by
comparing the data with a running mean over 10 phase steps. The running mean
has, thus, a reduced $\chi^2$ of $1$. Since the best fit to artificial data
with low synthetic noise levels does better than the running mean, we obtain
$\chi^2/n<1$ for these light curves.

The entropy $S$ of the images is always close to 1. An ideal entropic image
would have $S=1$, the actual values of $S=0.94\dots 0.98$ come very close to
that. One would not expect $S=1$, since that would imply that there is no
variation in the image, i.e., all surface elements have the same
brightness. The trend to higher $S$ with lower noise is a result of the
weighting of $S$ with respect to $\chi^2/n$ in the one-dimensional quality
parameter ${\cal Q}$ (Eq. \ref{e_quality}). For the test calculations, we have
chosen $\lambda=1$ constant for all noise levels. This emphasizes the
reduction of $\chi^2$ for low noise levels and the smoothing for high noise
levels.

Our tests show that we are able to detect structures in the brightness
distribution on the accretion stream with a size of $\approx 1\dots2 R_1$ from
data with phase coverage, phase resolution and noise level similar to that of
the HST data set for UZ\;For (see section \ref{s_data}).

\subsection{Full-orbit light curve}

\begin{figure}[tb]
\begin{center}
\resizebox{7cm}{!}{\includegraphics{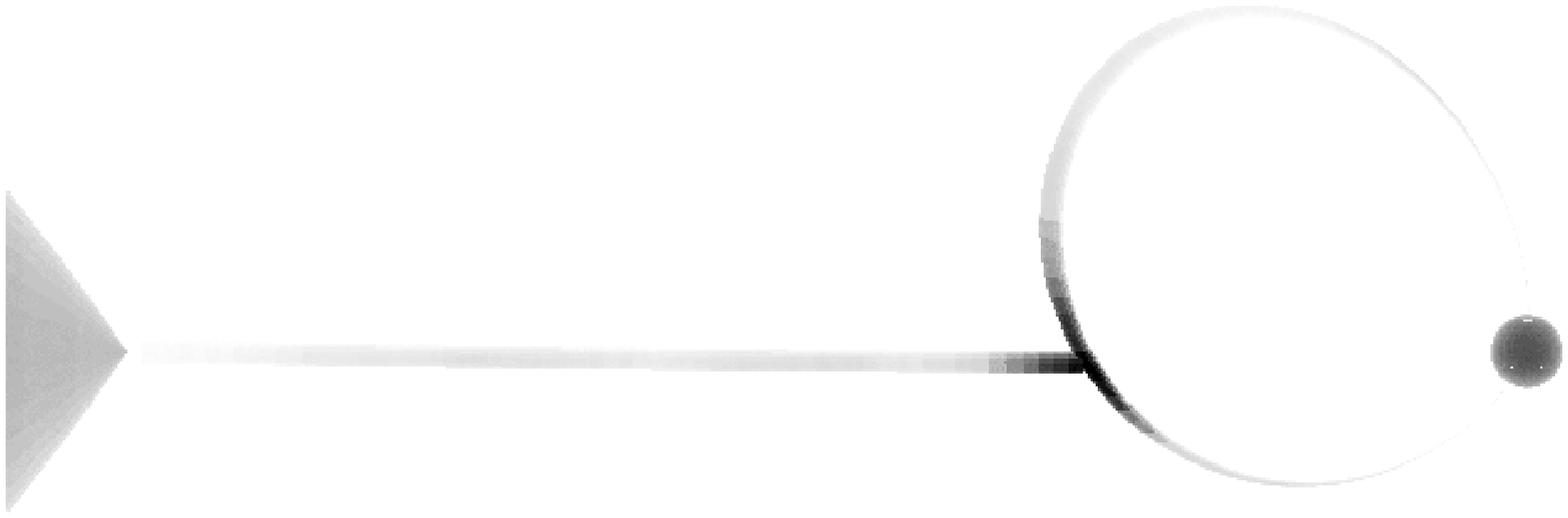}}

\bigskip
\resizebox{8.8cm}{!}{\includegraphics{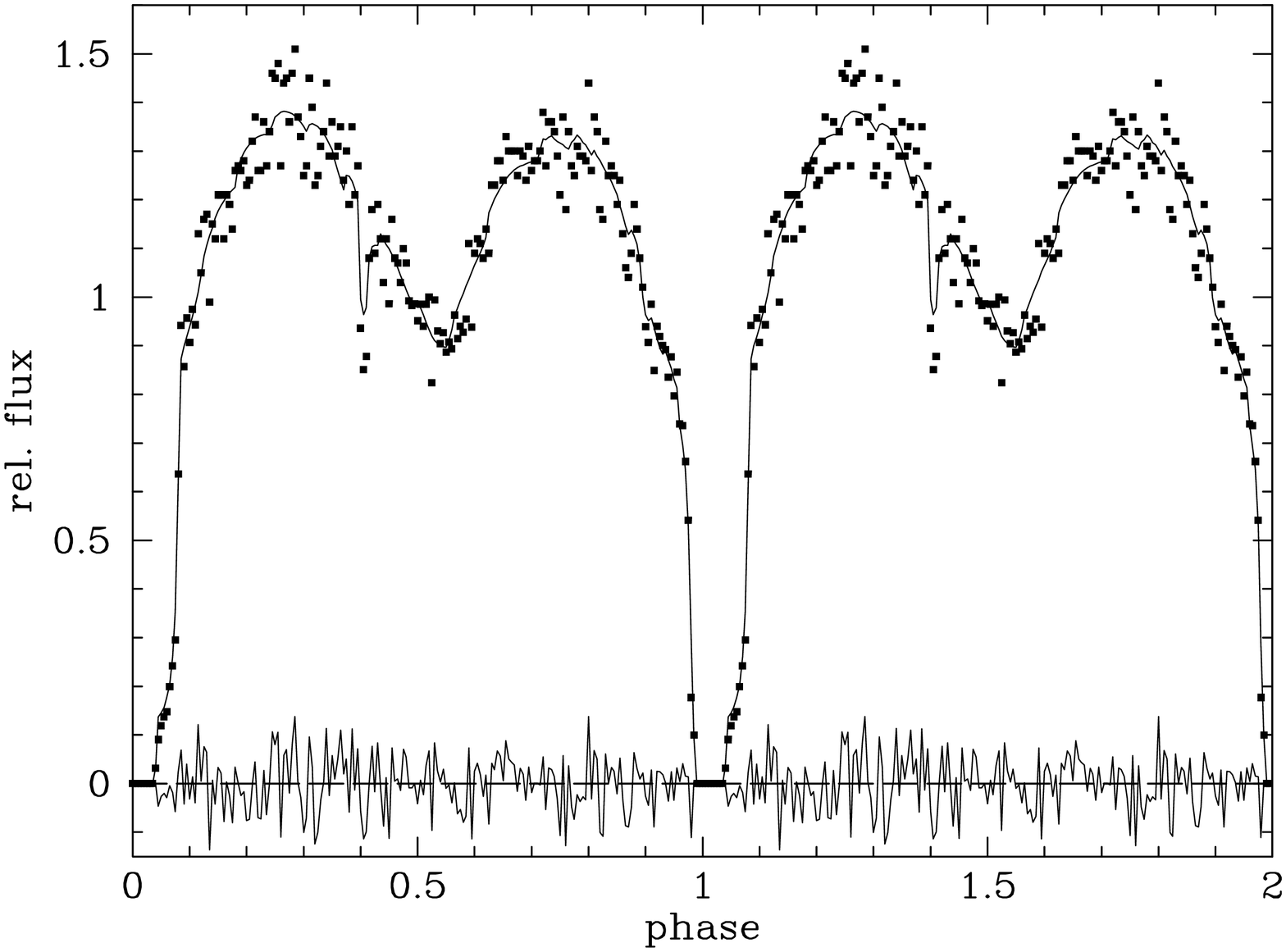}}
\end{center}
\caption{Top: Full-orbit reconstruction of the `point'-brightness
distribution. Compare to Fig. \ref{f_map_point}, top. Bottom: Full-orbit
synthetic light curve and best-fit light curve. For clarity, two full orbits
are shown.}
\label{f_full}

\end{figure}

To test the capabilities of our algorithm for data with wider phase coverage,
we fit a synthetic light curve covering the whole binary orbit,
computed using the same input map and geometry as in
Sect. \ref{ObrnSR}. We choose a phase resolution
of 0.005 for the simulated data, corresponding to a 30\,sec time resolution, and S/N=10.
The result of this fit is shown in Fig. \ref{f_full}. Obviously, the
additional phase information helps in producing a reliable reconstruction of
the initial intensity distribution, which one can see by comparing the map in
Fig. \ref{f_full} with the $\mbox{S/N}=10$-map in Fig. \ref{f_map_point}. The
full-orbit light-curve has less phase steps than the ingress-only-map, but
shows the same quality of the reconstruction. On the other hand, the
similarity between the two results allows us to conclude that we can rely on
the reconstructions of our algorithm even 
if only ingress data is available, as will be the case for the HST
archive data of UZ\;For which we use in the following.

The synthetic light curve over the full orbit shows various
eclipse and projection features, which are described in detail by Kube
et al. \cite*{KGB98a}.

\section{Application: UZ For}

\subsection{System Geometry of UZ For}

UZ For has been identified as a polar in 1988 \cite{BS88,BTS88,OGA88}.
Cyclotron radiation from a region with $B=53\,\mbox{MG}$ has been reported by
Schwope et al. \cite*{SBT90} and Rousseau et al. \cite*{RFB96}. 
The first mass estimates for the WD were rather high,
$M_1=1.09\pm0.01M_{\sun}$ and $M_1>0.93M_{\sun}$ \cite{HKL88,BTS88}, but
Bailey and Cropper \cite*{BC91} and Schwope et al. \cite*{SMB97} derived
significantly smaller masses, $0.61M_{\sun}<M_1<0.79M_{\sun}$ and
$M_1=0.75M_{\sun}$, respectively.
We use reliable system parameters from Bailey and Cropper \cite*{BC91},
$q=M_1/M_2=5$, $M=M_1+M_2=0.85\,M_\odot$, $i=81^\circ$,
$P=126.5\mathrm{\,min}$. The optical light curve in Bailey \cite*{B95} shows
two eclipse steps which are interpreted as the signature of hot spots near
both magnetic poles of the WD, consecutively dissapearing behind the limb of
the secondary star. From that light curve we measure the timing of the eclipse
events with an accuracy of $\Delta\Phi=5\cdot10^{-4}$. The ingress of the spot
on the lower hemisphere occurs at $\Phi=0.9685$, its egress at $\Phi=1.0310$.
For the spot on the upper hemisphere, ingress is at $\Phi=0.9725$ and egress
at $\Phi=1.0260$.

To describe the spatial position of the dipole field line along which
the matter is accreted, three angles are needed: The colatitude or
tilt of the dipole axis $\beta$, the longitude of the dipole axis
$\Psi$, and the longitude of the stagnation region $\Psi_S$. Here,
longitude is the angle between the secondary star and the
respective point as seen from the centre of the white
dwarf.  With our choice of $\beta$, $\Psi,$ and $\Psi_S$, we can
reproduce the ingress and egress of the two hot spots as well as the
dip at phase $\Phi=0.9$. A
summary of the main system parameters used in this analysis is given in
Table \ref{t_general_parameters}.

\begin{table}

\caption{System geometry of UZ Fornacis}
\label{t_general_parameters}
\begin{flushleft}
\begin{tabular}{ll}
\hline\noalign{\smallskip}
mass ratio&$Q=M_1/M_2=5$\\
total mass&$M=M_1+M_2=0.85M_{\sun}$\\
orbital period&$P=126.526\,229\,\mbox{min}$\\
orbital separation&$a=5.49\times10^{10}\,\mbox{cm}$\\
inclination&$i=81\degr$\\
radius of WD&$R_1=7.53\times10^8\,\mbox{cm}$\\
\noalign{\smallskip}\hline\noalign{\smallskip}
dipole tilt&$\beta=12\degr$\\
dipole azimuth&$\Psi=5\degr$\\
azimuth of stagnation region&$\Psi_S=34\degr$\\
`radius' of ballistic stream&$r_S=5\times10^8\,\mbox{cm}$\\
\noalign{\smallskip}\hline
\end{tabular}
\end{flushleft}
\end{table}

Having the correct geometry of the accretion stream is crucial for
generating the correct reconstruction of the emission regions. As we have
shown in \cite{KGB98a}
especially the geometry of the dipole stream is sensitive to changes in
$\Psi$, $\Psi_S$, and $\beta$. For UZ\;For, the geometry of the dipole stream
is relatively well constrained from the observed ingress and egress of both
hot spots on the white dwarf \cite{B95}. Fitting the UZ\;For light curves
system geometries that differ within the estimated error range of less than
five degrees
does, however, not significantly affect our results. This situation is
different if the errors in the geometry parameters are larger than only a few
degrees.

\subsection{Observational Data}
\label{s_data}

UZ\;For was observed with HST on June 11, 1992. A detailed description of
the data is given by Stockman and Schmidt \cite*{SS96}. We summarize here
only the relevant points. 

\begin{figure}
\includegraphics[width=8.8cm]{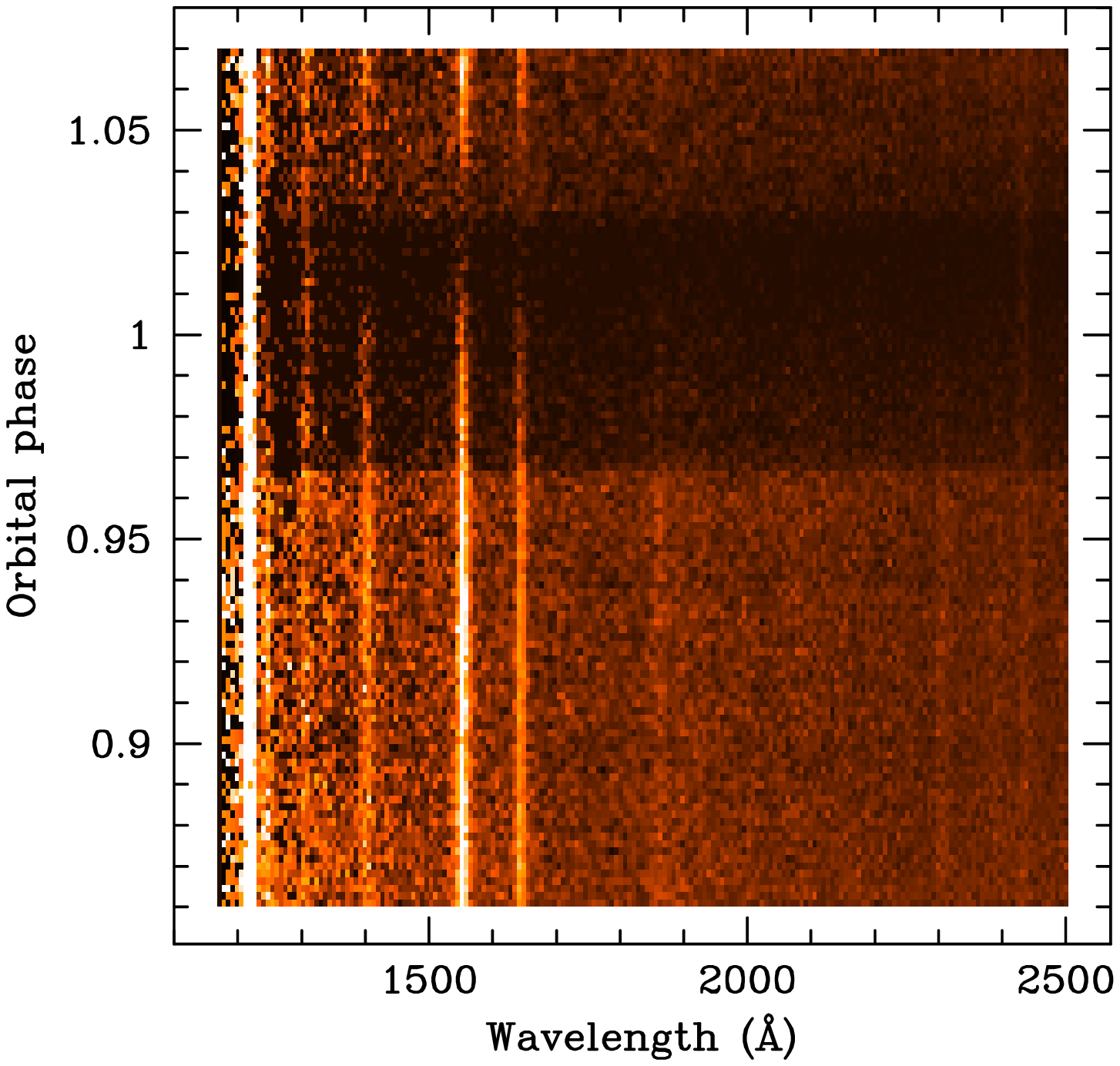}
\caption{Trailed spectrum of UZ\;For, both observed orbits added and
rebinned. The figure clearly shows the abrupt ingress and egress of
the continuum source and the more gradual eclipse of the emission line
source. It also shows a faint dip in the continuum and in the lines at
$\Phi=0.90$ which occurs when the magnetically funneled section of the
accretion stream crosses the line of sight to the white dwarf.}
\label{f_trailed}
\end{figure}

Fast FOS/G160L spectroscopy with a time resolution of 1.6914\,s was
obtained, covering two entire eclipses in the phase interval
$\Phi=0.87\dots1.07$. The two eclipses were observed starting at
05:05:33~UTC (`orbit 1') and 11:25:39~UTC (`orbit 2'). The spectra
cover the range $1180\dots2500$\,\AA\ with a FWHM resolution of
$\approx7$\,\AA. The mid-exposure times of the individual spectra were
converted into binary orbital phases using the ephemeris of Warren et
al. \cite*{WSV95}. The average trailed spectrum is shown in
Fig.~\ref{f_trailed}.

\begin{figure*}
\includegraphics[width=8.8cm]{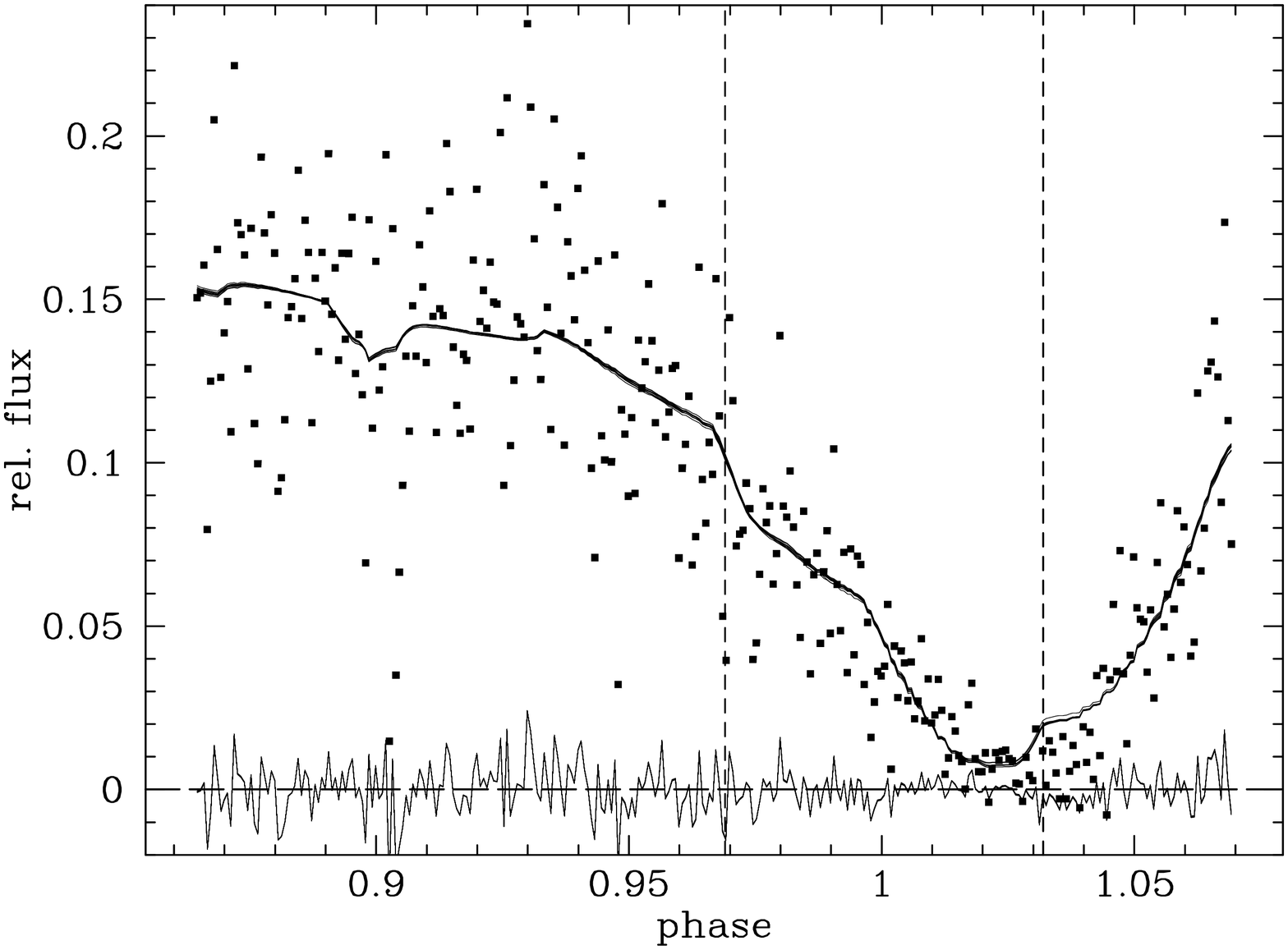}
\includegraphics[width=8.8cm]{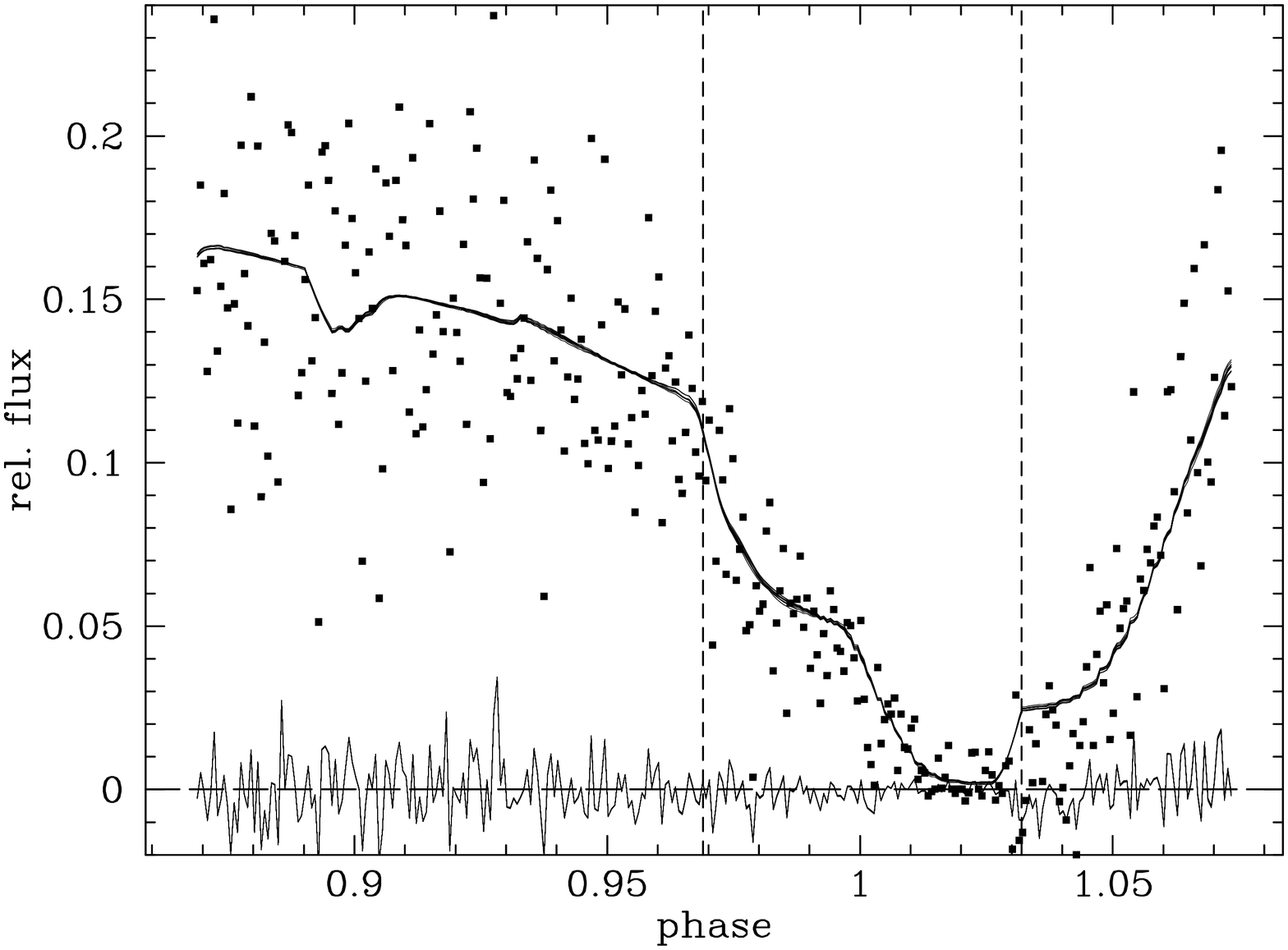}
\caption{Extracted \ion{C}{iv} light curves and best fits. Left: orbit 1,
right: orbit 2. In each panel, ten light curves from different fit runs are
overplotted (hard to recognize) to show the stability of the fit. The
residuals are scaled down by a factor of 4 for clarity. The vertical
dashed lines mark the ingress and egress of the white dwarf and, hence,
approximately the begin of the ingress and egress of the magnetically funneled
accretion stream.}
\label{f_lcs}
\end{figure*}

In order to obtain a light curve dominated by the accretion stream, we
extracted the continuum subtracted \Line{C}{iv}{1550} emission from the
trailed spectrum.  The resulting light curves are shown in Fig.~\ref{f_lcs}
for both orbits separately.  To reduce the noise to a bearable amount, the
light curves were rebinned to $5.07\,\mbox{s}$ resulting in a phase
resolution of $\Delta\Phi=6.7\times10^{-4}$.

We note that the \ion{C}{iv} light curve may be contaminated by emission from
the heated side of the secondary star.  HST/GHRS observations of AM\;Her,
which resolve the broad component originating in the stream and the narrow
component originating on the secondary, show that the contribution of the
narrow component to the total flux of \ion{C}{iv} is unlikely to be larger
than $10\dots15$\,\% \cite{GHB98}. Furthermore, during the phase interval
covered in the HST observations of UZ\;For, the irradiated hemisphere of the
secondary is (almost) completely self-eclipsed, so that its \ion{C}{iv}
emission is minimized.

\subsection{Results}
\label{Results}

\begin{figure*}
\begin{center}
\includegraphics[width=8.8cm]{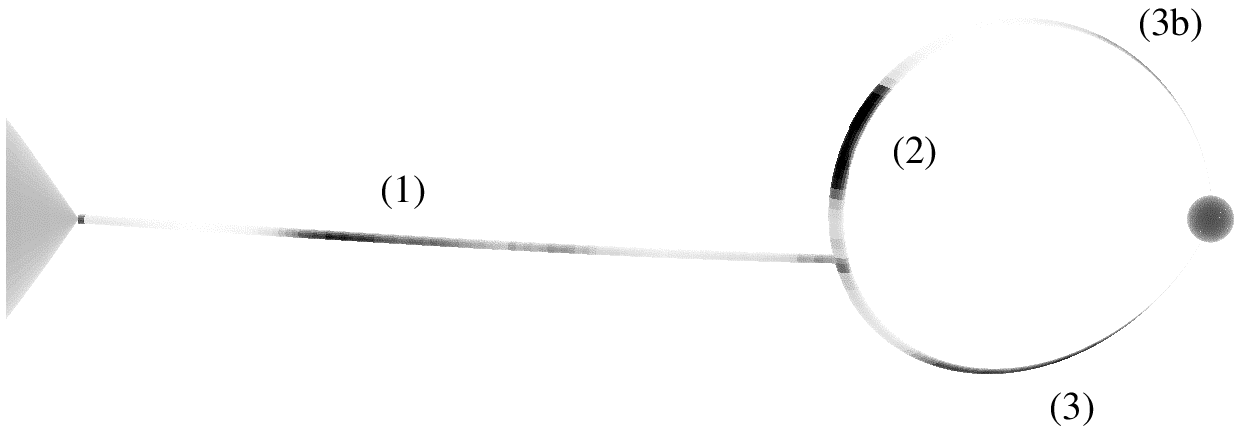}
\includegraphics[width=8.8cm]{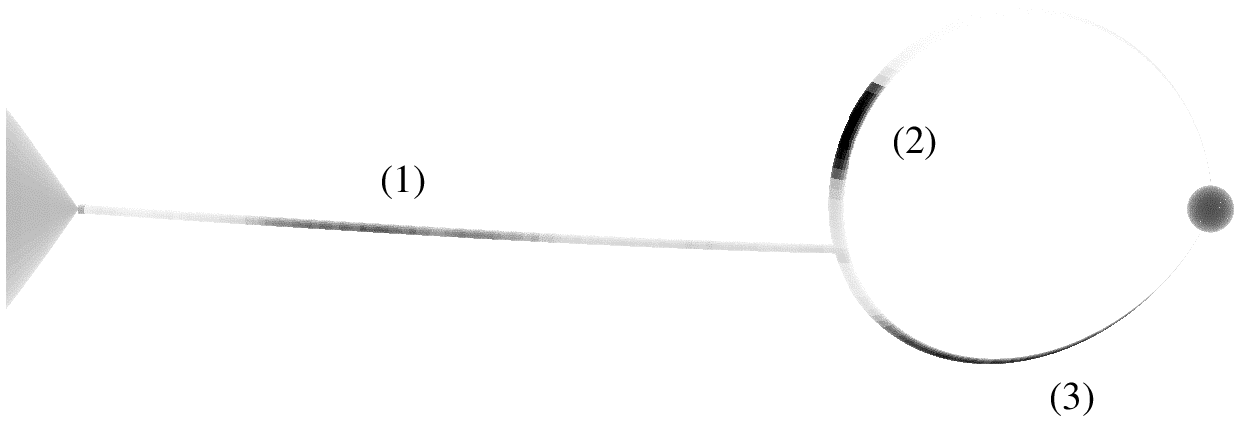}
\end{center}

\caption{Resulting intensity maps of the accretion stream in
UZ\;For. Left: Orbit 1, right: Orbit 2. Bright regions are
printed in black, dark regions in white.}
\label{f_result_map}
\end{figure*}

\begin{figure*}
\begin{center}
\includegraphics[width=8.8cm]{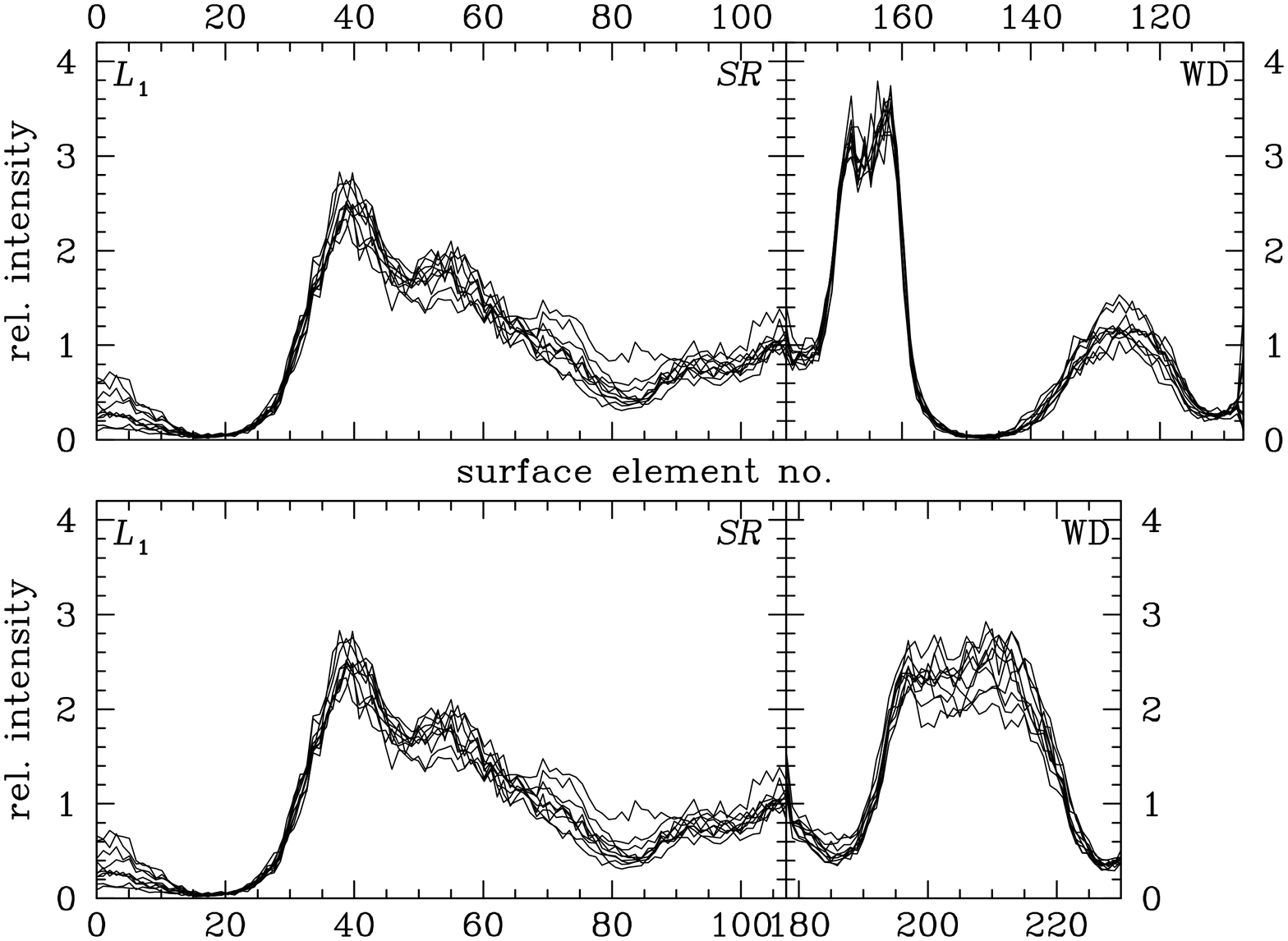}
\includegraphics[width=8.8cm]{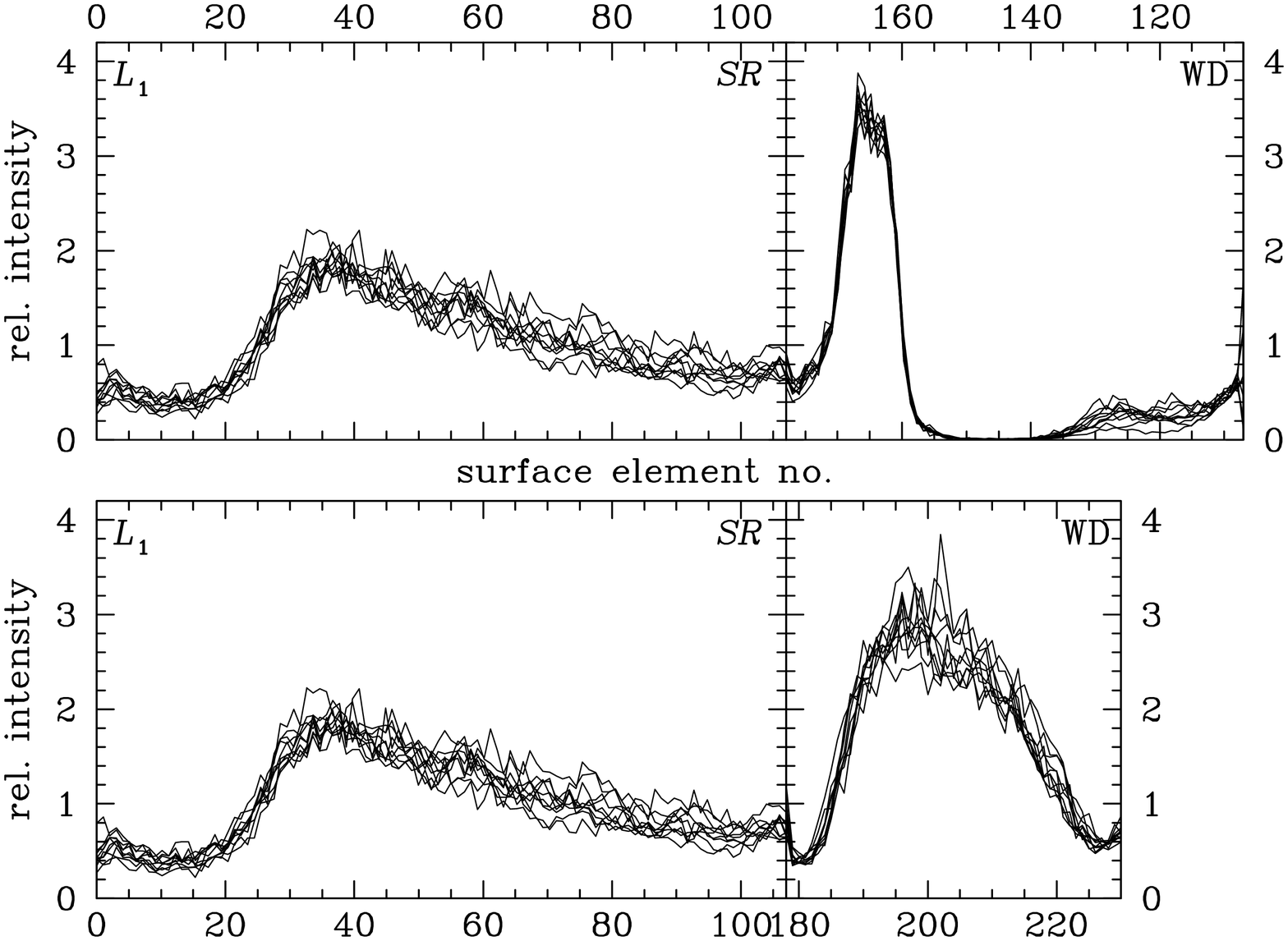}
\end{center}

\caption{Resulting intensity distributions of the accretion stream in
UZ\;For. Left: Orbit 1, right: Orbit 2. The results of 10 individual fit runs
are overplotted in one graph to show the consistency of the fits. For an
explanation of the plots see Fig.~\ref{f_plot_point}.}
\label{f_result_map_plot}
\end{figure*}

The light curves show small, but significant differences for the two orbits
(Fig. \ref{f_lcs}). In orbit 1, the dip at $\Phi=0.9$ is slightly
deeper than in orbit 2. However, this very small feature has only a
marginal effect on the results. The dip is well known from X-Ray and
EUV observations \cite{WSV95} and 
has been observed to move 
in phase between $\Phi=0.88$ and
$0.92$ on timescales of months \cite{SH98}. The ingress of the accretion
stream into eclipse is much smoother in orbit 1 than in orbit 2, where an
intermediate brightness level around $\Phi=0.98$ with a flatter slope is seen.
The \ion{C}{iv} intensity maps resulting from our fits are shown in
Fig.~\ref{f_result_map} for each observation interval separately. In
Fig.~\ref{f_result_map_plot}, we show the relative intensity distributions of
10 fit runs for each orbit, proving that our algorithm finds the same result
(except for noise) for each run. In Fig.~\ref{f_result_map}, the resulting map
from one arbitrary fit is shown.

The brightness maps for the two orbits show common features and
differences. Common in both reconstructed maps are the bright regions (1) on
the ballistic stream, (2) on the dipole stream above the orbital plane, and
(3) on the dipole stream below the orbital plane. In orbit 1, there is an
additional bright region on the northern dipole stream which appears as a
mirror image of region 3. We denominate it 3b. The difference between both
maps is found in the presence/absence of region 3b, and in the different
sharpness of region 1, which is much brighter and more peaked in orbit 1 than
in orbit 2.

Remarkable is that we do not find a bright region at the coupling region $SR$,
where one would expect dissipative heating when the matter rams into the
magnetic field and is decelerated. We will comment on this result in
Sect.~\ref{saesp}.

The sharp upper border of region 2 has to be discussed separately: As one can
see from the data, the flux of the \Line{C}{iv}{1550} emission ceases
completely in the phase interval $\Phi\approx0.01\dots0.03$. Hence, all parts
of the accretion stream which are not eclipsed during this phase interval can
not emit light in \Line{C}{iv}{1550}. For the assumed geometry of UZ\;For,
parts of the northern dipole stream remain visible throughout the
eclipse. Thus, the sharp limitation of region 2 marks the border between those
surface elements which are always visible and those which dissappear behind
the secondary star. Uncertainties in the geometry could affect the location of
the northern boundary of region 2, but should not change the general result,
namely that there is emission \emph{above} the orbital plane that accounts for
a large part of the total stream emission in \Line{C}{iv}{1550}.

Region 3b has to be understood as an artifact: 
During orbit 1, the
observed flux level at maximum emission line eclipse
($\Phi=0.01\dots0.03$)
does not drop to zero.
Hence, our algorithm places intensity on the surface elements of the
accretion stream which are still visible at that phase.
Apparently, the evolution strategy tends to place these residual
emission not uniformly on all the visible surface elements but on
those closer to the WD, which leads to an intensity pattern that
resembles the more intense region on the southern side of the dipole
stream.

\begin{figure}[t]
\begin{center}
\includegraphics[width=8.8cm]{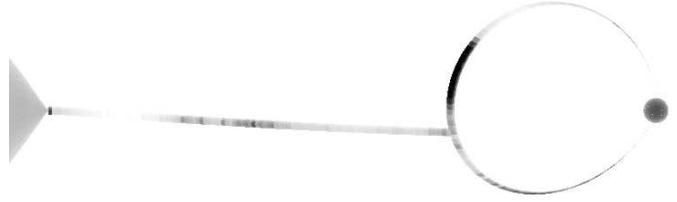}
\end{center}
\caption{Intensity map of the accretion stream for orbit 1. The light curve
used to generate this map was changed from the observed so that for each phase
point the flux was randomly modified with a gaussian with $\sigma$ as
described in Sect. \ref{Tbrotbs}. Compare with Fig. \ref{f_result_map} and see
text.}
\label{f_map_rand}
\end{figure}

To underpin the fact that 
regions 1, 2, and 3 in our map are
real features and not just regions which result by random
fluctuations in the data, we test what happens to the reconstruction
if the input light curve is changend. For the calculation which
results in the map shown in Fig. \ref{f_map_rand}, we generated a
modified light curve from the data for orbit 1. For each phase step,
we modified the flux, so that $F'(\Phi)=F(\Phi)+1/2\cdot\sigma(\Phi)\cdot G$
is the new value. $\sigma(\Phi)$ is the local standard deviation as
defined in Sect. \ref{Tbrotbs}, $G$ are gaussian-distributed random
values. Since the map from the light curve $F'(\Phi)$ does not show
significant differences from the map corresponding to the original
data $F(\Phi)$ (Fig. \ref{f_result_map}), we conclude that the
features 1, 2, and 3 are real.

\section{Discussion}
We have, for the first time, mapped the accretion stream in a polar in
the light of a high-excitation ultraviolet line with a complete 3d
model of an optically thick stream. We have found three different
bright regions on the stream, but no strong emission at the stagnation
point of the ballistic stream. In the following we will discuss the
physical processes which may lead to an emission structure like the
one observed.

\subsection{Emission of the ballistic stream} 

As mentioned in Sect. \ref{s_Method}, single-particle trajectories with
different inital directions diverge after the injection at $L_1$, but converge
again at a point approximately one third of the way between $L_1$ and the
stagnation region. This is where we find emission in the line of
\Line{C}{iv}{1550}. Possibly the kinetic properties of the stream lead to a
compression of the accreted matter, resulting in localized heating. After
the convergence point, the single-particle trajectories diverge slowly and
follow a nearly straight path without any further stricture. Hydrodynamical
modelling of the ballistic part of the stream is required to substantiate this
hypothesis.

\subsection{Absence of emission at the stagnation point}
\label{saesp}
In the classical model of polars, it is assumed that the ballistically
infalling matter couples onto the magnetic field in the stagnation region with
associated dissipation of kinetic energy (e.g. Hameury et al.,
1986)\nocite{HKL86}. Thus, one would expect a bright region near
$SR$.
The absence of \Line{C}{iv}{1550} emision in the stagnation region could be
due to the fact that there is no strong heating in the coupling
region. Dissipation near $SR$ can be avoided if the material is continously
stripped from the ballistic stream and couples softly onto the field lines, as
proposed by Heerlein et al. \cite*{HHS98} for HU\;Aquarii.

Another possibility is that the matter is decelerated near $SR$,
resulting in an increase in the density. This may result in an
increase of the continuum optical depth, and, therefore, in a decrease
of the \Line{C}{iv}{1550} equivalent width.

\subsection{Emission of the dipole stream}

On the dipole section of the stream, we find two generally different emission
regions: The bright and small region above the stagnation point and
the broader regions near the accretion poles of the white dwarf.

\begin{figure}
\begin{center}
\includegraphics[width=6cm]{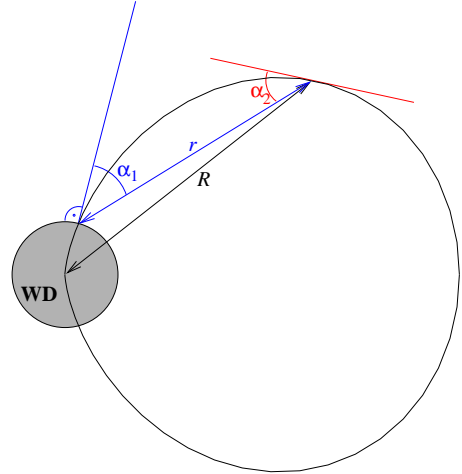}
\end{center}
\caption{Dipole field line configuration. See Sect.~\ref{s_disc_accspot}.}
\label{f_dipol}
\end{figure}

{\em Near the accretion spots:}
\label{s_disc_accspot}
On the magnetically funneled stream, we find one 
region of line emission (3) which we assume to be due to
photoionization by 
high-energy radiation from the accretion spot. 
The mirror region 3b is an artifact which is created by the mapping
algorithm to account for the non-zero flux level in orbit 1 in the
phase interval $\Phi=0.01\dots0.03$ (see Sect. \ref{Results}).
Even though the distribution of \ion{C}{iv} emission on  the accretion
stream does not reflect the irradiation pattern in a straigthforward way,
the presence of \ion{C}{iv} emission at a certain location of the
stream requires that this point is irradiated, if there is no other mechanism
creating \ion{C}{iv} line emission. In Fig.~\ref{f_dipol},
we define the angles $\alpha_1$ and $\alpha_2$ and the distances $r$
and $R$. Assuming a point-like emission region at the accretion spot,
the absorbed energy flux per unit area caused by illumiation
from the accretion regions varies as $\cos\alpha_1\sin\alpha_2
R^{3/2}/{r^2}$.
The structure of this equation shows that the illumination of the accretion
stream is at maximum somewhere between the accretion region and the point with
$\alpha_1=0$. This corresponds to the brightness distribution which we derive
from the data.

We suggest the following interpretation of the emission region
\emph{above the stagnation point}: Independent
of whether the \ion{C}{iv} emission in this region is due to photoionozation
or collisional excitation, a higher density than in other sections of the
stream is required.  Matter which couples in $SR$ onto the field lines has
enough kinetic energy to initially rise northward from the orbital plane
against the gravitational potential of the white dwarf. If the
kinetic energy is not sufficient to overcome the potential summit on the field
line, the matter will
stagnate 
and eventually fall back towards the orbital
plane, where it collides with further material flowing up. This may lead to
shock heating with subsequent emission of
\Line{C}{iv}{1550}. Alternatively, the photoinization of the region
of increased density may suffice to create the emission peak.  Yet another
possibility is that the matter is heated by cyclotron radiation from the
accretion column which emerges preferentially in direction perpendicular to
the field direction. Crude estimates show, however, that the energy of the
cyclotron emission does not suffice to produce such a prominent feature as
observed. 
A detailed understanding of the emission processes in the
accretion stream involves a high level of magnetohydrodynamical
simulations and radiation transfer
calculations,
which is, clearly, beyond the
scope of this paper.

\subsection{Conclusion}

We have successfully applied our new 3d eclipse mapping method to
UZ\;Fornacis. In subsequent research we will allow additional degrees of
freedom in the mapping process, using data sets with higher S/N and covering a
larger phase interval.

Our attempts to image the accretion stream in polars should help in
understanding the physical conditions in the stream, such as density and
temperature. By comparison to hydrodynamical stream simulations, we will take
a step towards the complete understanding of accretion physics in polars.

\begin{acknowledgements}
We thank Hans-Christoph Thomas for discussions on the geometry of the
accretion stream, Andreas Fischer for general comments on this work, Andr\'e
Van Teeseling for ideas to test the algorithm and an anonymous referee for
the detailed and helpful comments. Part of this work was funded by the DLR
under contract 50\,OR\,99\,03\,6.

\end{acknowledgements}

\end{document}